\documentstyle[11pt,epsf]{article}

\newcommand{\sect}[1]{ \section{#1} \setcounter{equation}{0} }

\newcommand{\kslash}{k \! \! \! /}

\newcommand{\Dslash}{D \! \! \! \! /}

\newcommand{\half}{\mbox{\small{$\frac{1}{2}$}}}

\newcommand{\Nf}{N_{\!f}} 
\newcommand{\MSbar}{\overline{\mbox{MS}}} 

\begin{document}
\title{Anomalous dimensions of operators in polarized deep inelastic 
scattering at $O(1/N_{\! f})$}  
\author{J.A. Gracey, \\ Department of Mathematical Sciences, \\ University of 
Liverpool, \\ P.O. Box 147, \\ Liverpool, \\ L69 3BX, \\ United Kingdom.}
\date{} 
\maketitle
\vspace{5cm}
\noindent
{\bf Abstract.} Critical exponents are computed for a variety of twist-$2$ 
composite operators, which occur in polarized and unpolarized deep inelastic
scattering, at leading order in the $1/\Nf$ expansion. The resulting 
$d$-dimensional expressions, which depend on the moment of the operator, are in 
agreement with recent explicit two and three loop perturbative calculations. An
interesting aspect of the critical point approach which is used, is that the 
anomalous dimensions of the flavour singlet eigenoperators, which diagonalize 
the perturbative mixing matrix, are computed directly. We also elucidate the 
treatment of $\gamma^5$ at the fixed point which is important in simplifying
the calculation for polarized operators. Finally, the anomalous dimension of 
the singlet axial current is determined at $O(1/\Nf)$ by considering the 
renormalization of the anomaly in operator form.  

\vspace{-19cm} 
\hspace{13.5cm} 
{\bf LTH 384} 

\newpage

\sect{Introduction.} 
Our understanding of the structure of nucleons is derived primarily from 
experiments where they are bombarded by other nucleons or electrons at high 
energies. These deeply inelastic processes are, in general, well understood in 
most instances. Current activity, however, centres on examining polarized 
reactions due, for example, to the discrepancy observed in results for the spin
of the proton and theoretical predictions by the EMC collaboration, \cite{1}. 
Consequently in order to make accurate statements about the data current 
theoretical interest has focussed on carrying out higher order perturbative 
calculations in the underlying field theory, quantum chromodynamics, (QCD). As
this theory is asymptotically free at high energies, \cite{2}, the coupling 
constant is sufficiently small so that perturbative calculations give a good 
description of the deep inelastic phenomenology. Indeed unpolarized scattering 
is well understood with one and two loop results available for the anomalous 
dimensions of the twist-$2$ flavour non-singlet and singlet operators which 
arise in the operator product expansion, [3-5]. Moreover, the Dokshitzer, 
Gribov, Lipatov, Altarelli, Parisi, (DGLAP), splitting functions, \cite{6}, 
which are a measure of the probability that a constituent parton fragments into
other partons are known to the same accuracy, \cite{3,5}. The moments of the 
scattering amplitudes have also been studied. More recently the three loop 
structure has been obtained exactly, for low moments, through huge impressive 
analytic computations, both for non-singlet and singlet cases, \cite{7}. 

The situation for polarized scattering is less well established. The one loop 
anomalous dimensions for the corresponding twist-$2$ (and $3$) operators were 
computed by Ahmed and Ross in \cite{8}. However, only recently has the two
loop structure been determined for the flavour singlet operators, \cite{9}. 
(The non-singlet polarized dimensions are equivalent to the non-singlet 
unpolarized case.) This was checked by Vogelsang in \cite{10,11} by 
calculating the splitting functions themselves and then comparing with \cite{9}
by taking the inverse Mellin transform. In this the moment $n$ of the operator
has the conjugate variable $x$ which is the momentum fraction of the parton in
the nucleon. These results have been important for the next to leading order 
evolution of the structure functions to low $x$ and $Q^2$ regions, \cite{12}.
(For review articles see \cite{13}.) Crucial in this exercise is the 
dependence of the results on the moment of the operators. 

To go beyond this two loop picture would require a great amount of 
computation based, for example, on the unpolarized results of \cite{7}. One 
way of improving our knowledge would be to compute the appropriate quantities 
using another approximation. For example, the large $\Nf$ expansion, where 
$\Nf$ is the number of quark flavours, has been used to obtain the leading 
order coefficients of the anomalous dimensions of the twist-$2$ unpolarized 
non-singlet operators to all orders in the perturbative coupling constant, 
\cite{14}. The resulting analytic function of $n$ provided a useful check on 
the exact low moment non-singlet three loop calculation of \cite{7}. Briefly 
the method involved studying the scaling behaviour of the operator at the 
non-trivial fixed point in $d$-dimensional QCD. With that success and the 
current interest in polarized physics required for exploring new $x$ 
r\'{e}gimes it is appropriate to apply the large $\Nf$ analysis to study the 
dimensions of the underlying (twist-$2$) operators. Aside from providing 
$n$-dependent results and getting a flavour of the structure beyond two loops, 
it will give at least another partial check on the recent results of [9-11]. In
particular we will determine the critical exponents at $O(1/\Nf)$ which encode 
the all orders coefficients of the twist-$2$ polarized singlet operators. As a 
prelude we need to study the unpolarized singlet case which will extend the 
result of \cite{14}. 

The paper is organised as follows. The basic formalism and notation is 
introduced in section 2. We review previous work in section 3, including the 
technical details of the computation of singlet unpolarized operators anomalous
dimensions. As the treatment of four dimensional objects, such as $\gamma^5$, 
is needed we review previous $1/\Nf$ work involving this in section 4. The 
remaining sections 5 and 6 are devoted to the application of the results of  
earlier sections to polarized operators. In particular the latter section 
centers on the treatment of the singlet axial current which is not conserved
due to the chiral anomaly. Future work and perspectives are discussed in 
section 7. An appendix gives details of the relation of the $1/\Nf$ exponent
results to the DGLAP splitting functions.  

\sect{Background.} 

To begin with we review several of the more field theoretic aspects of the 
formalism including the role of the critical renormalization group. First, we 
recall that critical exponents are fundamental quantities. In experiments and
condensed matter problems dealing with phase transitions they completely 
characterize the physics. In order to describe such phenomena one determines 
estimates of the exponents by calculating, for example, in the underlying 
quantum field theory describing the transition. In practice this means 
carrying out a perturbative calculation of the renormalization constants of 
the theory, in some renormalization scheme. This information is then encoded
in the corresponding renormalization group functions such as the anomalous 
dimensions of the field or the mass. The critical coupling, $g_c$, is 
subsequently determined from the $\beta$-function of the theory. It is defined
to be a non-trivial zero of $\beta(g)$. The appropriate critical exponents are
then found by evaluating the anomalous dimensions at the critical coupling. In
practice provided enough terms of the series have been computed, relatively 
accurate numerical estimates can be obtained. (Useful background material can 
be found in, for example, \cite{15}.)  

For the present problem we will examine a fixed point in QCD in $d$-dimensions 
and obtain the critical exponents as a function of $d$ and $\Nf$ which 
characterize the transition. As indicated these correspond to the 
renormalization group equation, (RGE), functions evaluated at $g_c$. Therefore 
provided the location of $g_c$ is known in some approximation like $1/\Nf$ one 
can decode the information contained in the exponent and determine the 
coefficients of the RGE functions for non-critical values of the coupling. In 
this large $\Nf$ method it turns out that the structure of $\beta(g)$ is such 
that for leading order calculations in $1/\Nf$ only knowledge of the one loop 
coefficient is required, \cite{16}. We illustrate these remarks with a general 
example, which will set notation for later sections. As we are interested in 
the coefficients in the series of an RGE function we denote such a function by 
$\gamma(g)$ and define its expansion to be, with explicit $\Nf$ dependence,  
\begin{equation} 
\gamma(g) ~=~ a_1g ~+~ (a_2\Nf + b_1)g^2 ~+~ (a_3{\Nf}^2 + b_2\Nf + c_1)g^3 
{}~+~ O(g^4) 
\end{equation} 
where the obvious definition for the $O(g^4)$ term is understood. The 
coefficients $\{ a_i$, $b_i$, $c_i$, $\ldots \}$ can of course be functions of
other parameters such as colour group Casimirs or $n$. To evaluate (2.1) at a 
fixed point we take the general structure of the $\beta$-function to be, in 
$d$-dimensions,  
\begin{equation} 
\beta(g) ~=~ (d-4)g ~+~ A\Nf g^2 ~+~ (B\Nf + C)g^3 ~+~ O(g^4) 
\end{equation} 
Setting $d$ $=$ $4$ $-$ $2\epsilon$, then there is a non-trivial fixed point, 
$g_c$, at 
\begin{equation} 
g_c ~=~ \frac{2\epsilon}{A\Nf} ~+~ O \left( \frac{1}{\Nf^2} \right) 
\end{equation} 
So at $O(1/\Nf)$ 
\begin{equation} 
\eta ~\equiv~ \gamma(g_c) ~=~ \frac{1}{\Nf} \sum_{n=1}^\infty \frac{2^na_n 
\epsilon^n}{A^n} ~+~ O\left( \frac{1}{\Nf^2} \right) 
\end{equation} 
where $\eta$ is the corresponding critical exponent. Clearly at leading order 
in $1/\Nf$, when $\Nf$ is large, the coefficients $\{ a_i \}$ are accessed. To 
determine them one needs to compute $\eta$ directly in the $O(1/\Nf)$ expansion
in $d$-dimensions. This is the aim of the paper for the operators discussed 
earlier. 

Before reviewing that we make several parenthetical remarks. In assumimg a 
$\beta$-function of the form (2.2) we are restricting ourselves to a particular
class of theories which includes QED and QCD. If the two loop term of (2.2)
had been quadratic and not linear in $\Nf$ and likewise the three loop term
cubic in $\Nf$ and so on, then it would not be possible to determine a simple  
form for $g_c$ at leading order in $1/\Nf$. Instead an infinite number of 
terms of $\beta(g)$ would be required. This would imply a large $\Nf$ 
expansion would not be possible in that case. This is similar to the large 
$N_{\! c}$ expansion of QCD. Then the structure of the $\beta$-function has 
this nasty form and it is not easy to study QCD in a critical $1/N_{\! c}$ 
approach.  

The method to compute critical exponents corresponding to the anomalous 
dimensions of operators in powers of $1/\Nf$ is based on an impressive series 
of papers, [17-19]. In \cite{16,17} the $O(N)$ $\sigma$ model was considered
and the technique has been developed for fermion and gauge theories more
recently, \cite{20,21}. Essentially one studies the theory precisely at the
fixed point $g_c$ where there are several simplifications. First, at $g_c$ 
there is no mass in the problem so all propagators are massless. Second, the 
structure of the (full) propagators can be written down, in the approach to
criticality. Therefore in momentum space a fermion and gauge field will have
the respective propagators $\psi$ and $A_{\mu\nu}$ of the form, in the limit 
$k^2$ $\rightarrow$ $\infty$ \cite{16},  
\begin{equation} 
\psi(k) ~ \sim ~ \frac{A\kslash}{(k^2)^{\mu-\alpha}} ~~~,~~~ 
A_{\mu\nu}(k) ~ \sim ~ \frac{B}{(k^2)^{\mu-\beta}} \left[ \eta_{\mu\nu} ~-~ 
(1-b) \frac{k_\mu k_\nu}{k^2} \right] 
\end{equation}  
where $d$ $=$ $2\mu$, $b$ is the covariant gauge parameter and $A$ and $B$ are
amplitudes. The dimensions $\alpha$ and $\beta$ of the fields (in coordinate
space) comprise two pieces. For example, 
\begin{equation} 
\alpha ~=~ \mu ~-~ 1 ~+~ \half \eta 
\end{equation} 
where the first term is the canonical dimension of the fermion determined by
ensuring that the kinetic term in the action is dimensionless. The second term
is the critical exponent corresponding to the anomalous dimension of $\psi$ or
the wave function renormalization evaluated at $g_c$. It reflects the effect
radiative corrections have on the dimension of $\psi$. 

With (2.5) one can analyse any set of Feynman diagrams in the neighbourhood of
$g_c$ and determine the scaling behaviour of the integral. In particular one
can examine the $2$-point Schwinger Dyson equation at criticality to obtain a
representation of those equations. It turns out that one obtains a set of  
self-consistent equations which can be solved to determine $\eta$ analytically
as a function of $d$. Furthermore the approach is systematic in that 
$O(1/\Nf^2)$ corrections can be studied too. In \cite{19} this was extended
to $n$-point Green's function. If, for example, one considers the $3$-point
interaction then the exponent or the vertex anomalous dimension is found by
computing the (regularized) set of leading order integrals with (2.5). The 
residue of the simple pole of each graph contributes to the anomalous 
dimension. We will illustrate these remarks explicitly in the next section. 
However, we note that the regularization that is used is obtained by replacing
$\beta$ of (2.5) by $\beta$ $-$ $\Delta$. Here $\Delta$ is assumed to be small
like the $\epsilon$ used in dimensional regularization, \cite{19}.  

We conclude this section by recalling another feature of critical theory which
is important in analysing QCD in large $\Nf$. So far the above remarks have 
been completely general and summarize the approach taken in other models. 
Another common feature is that the theory that underlies a fixed point is not
necessarily the unique model describing the physics. More than one model can 
be used to determine the (measured) critical exponents. In this case such 
theories are said to be in the same universality class. From a field theoretic
point of view one can use this to simplify large $\Nf$ calculations. For 
example, the $O(N)$ $\sigma$ model and $\phi^4$ theory with an $O(N)$ 
symmetry are equivalent at the $d$-dimensional fixed point where the former is
defined in $2$ $+$ $\epsilon$ dimensions and the latter in $4$ $-$ $\epsilon$,
\cite{15}. The critical exponents computed in either are the same. For QCD 
there is also a similar equivalence which has been demonstrated by Hasenfratz
and Hasenfratz in \cite{12}. They showed that as $\Nf$ $\rightarrow$ $\infty$ 
QCD and a non-abelian version of the Thirring model are equivalent. The 
lagrangians of each are, for QCD, 
\begin{equation} 
L ~=~ i \bar{\psi}^{iI} ( \Dslash \psi)^{iI} ~-~ \frac{(G^a_{\mu\nu})^2}{4e^2}
\end{equation} 
and 
\begin{equation} 
L ~=~ i \bar{\psi}^{iI} ( \Dslash \psi)^{iI} ~-~ \frac{(A^a_\mu)^2}{2\lambda^2}
\end{equation} 
for the non-abelian Thirring model, (NATM), where $1$ $\leq$ $i$ $\leq$ $\Nf$,
$1$ $\leq$ $I$ $\leq$ $N_{\! c}$, $1$ $\leq$ $a$ $\leq$ $N^2_{\! c}$ $-$ $1$, 
$D_{\mu \, IJ}$ $=$ $\partial_\mu \delta_{IJ}$ $+$ $T^a_{IJ}A^a_\mu$, 
$G^a_{\mu\nu}$ $=$ $\partial_\mu A^a_\nu$ $-$ $\partial_\nu A^a_\mu$ $+$ 
$f^{abc}A^b_\mu A^c_\nu$, $\psi^{iI}$ is the quark field and $A^a_\mu$ is the
gluon field. The coupling constants of each lagrangian are $e$ and $\lambda$
and are dimensionless in $4$ and $2$ dimensions respectively. The auxiliary
spin-$1$ field of the NATM can be eliminated to produce a $4$-fermi interaction 
which is renormalizable in strictly two dimensions. Essentially at a fixed 
point it is the interactions which are important and which suggest (2.7) and 
(2.8) are equivalent. The quadratic terms serve only to define the canonical
dimensions of the fields and coupling constants as well as establishing various
scaling laws for the exponents. Although the NATM does not appear to contain 
the triple and quartic vertices typical of a Yang Mills theory, it was 
demonstrated in \cite{21} that as $\Nf$ $\rightarrow$ $\infty$ that they 
correctly emerge from the three and four point functions of (2.8) involving a 
quark loop. Therefore to compute fundamental critical exponents in QCD at 
$O(1/\Nf)$ it is sufficient to use the much simpler theory (2.8) to compute 
these as only one interaction arises. We will point out later where the 
contributions from $3$-gluon vertices would arise in the exponent calculation 
with respect to perturbation theory. Having defined the NATM in which we will 
calculate, we define $\beta$ as  
\begin{equation} 
\beta ~=~ 1 ~-~ \eta ~-~ \chi 
\end{equation} 
where $\chi$ is the quark gluon vertex anomalous dimension. Also in using a 
gauge field in a covariant gauge the ghost sector needs to be added to both
(2.7) and (2.8). However it turns out that at $O(1/\Nf)$ there are no 
contributions to any exponent we compute and we therefore omit them here, 
\cite{16}. 

Finally we state several earlier results which will be required. The 
$\beta$-function of QCD is \cite{2,23}, 
\begin{eqnarray}
\beta(g) &=& (d-4)g + \left[ \frac{2}{3}T(R)\Nf - \frac{11}{6}C_2(G) \right] 
g^2 \nonumber \\
&+& \left[ \frac{1}{2}C_2(R)T(R)\Nf + \frac{5}{6}C_2(G)T(R)\Nf
- \frac{17}{12}C^2_2(G) \right] g^3 \nonumber \\
&-& \left[ \frac{11}{72} C_2(R)T^2(R)\Nf^2 + \frac{79}{432} C_2(G) T^2(R)
\Nf^2 ~-~ \frac{205}{288}C_2(R)C_2(G)T(R)\Nf \right. \nonumber \\
&&+~ \left. \frac{1}{16} C^2_2(R) T(R) \Nf 
- \frac{1415}{864} C^2_2(G)T(R)\Nf + \frac{2857}{1728}C^3_2(G) 
\right] g^4 + O(g^5) 
\end{eqnarray} 
where the three loop term was given in \cite{24} and the colour group Casimirs
are defined as $\mbox{tr}(T^a T^b)$ $=$ $T(R)\delta^{ab}$, $T^a T^a$ $=$ 
$C_2(R)$ and $f^{acd} f^{bcd}$ $=$ $\delta^{ab}C_2(G)$. Although in our earlier
ansatz we omitted a constant term in the one loop coefficient its contribution 
to $g_c$ does not appear until $O(1/\Nf^2)$ and so 
\begin{equation}
g_c ~=~ \frac{3\epsilon}{T(R)\Nf} 
+ O \left( \frac{1}{\Nf^2} \right)
\end{equation}
Various basic exponents are known to $O(1/\Nf)$ and we note \cite{15} 
\begin{equation} 
\eta_1 ~=~ \frac{[(2\mu-1)(\mu-2)+\mu b] C_2(R)\eta^{\mbox{o}}_1} 
{(2\mu-1)(\mu-2)T(R)} 
\end{equation} 
where $\eta$ $=$ $\sum_{i=1}^\infty \eta_i(\epsilon)/\Nf^i$, 
$\eta^{\mbox{o}}_1$ $=$ 
$(2\mu-1)(\mu-2)\Gamma(2\mu)/[4\Gamma^2(\mu)\Gamma(\mu+1)\Gamma(2-\mu)]$ and 
\cite{16}  
\begin{equation} 
\chi_1 ~=~ -~ \frac{[(2\mu-1)(\mu-2)+\mu b] C_2(R)\eta^{\mbox{o}}_1} 
{(2\mu-1)(\mu-2)T(R)} ~-~ \frac{[(2\mu-1)+ b(\mu-1)] C_2(G)\eta^{\mbox{o}}_1} 
{2(2\mu-1)(\mu-2)T(R)} 
\end{equation} 
Throughout that paper we will work in an arbitrary covariant gauge. We note
that the physical operators which occur in the operator product expansion have 
gauge independent anomalous dimensions and by including a non-zero $b$ this 
will give us a minor check on the corresponding exponent calculations. The 
combination $z$ $=$ $A^2B$ arises too and 
\begin{equation} 
z_1 ~=~ \frac{\Gamma(\mu+1)\eta^{\mbox{o}}_1}{2(2\mu-1)(\mu-2)T(R)} 
\end{equation}  

\sect{Unpolarized operators.} 

We illustrate the large $\Nf$ technique by computing the critical exponent of
the simplest operator which arises in the operator product expansion. This is 
the twist-$2$ flavour non-singlet operator, \cite{25,3}, 
\begin{equation}
{\cal O}^{\mu_1 \ldots \mu_n}_{\mbox{\footnotesize{ns}},a} ~=~ 
i^{n-1} {\cal S} \bar{\psi}^I \gamma^{\mu_1} D^{\mu_2} \ldots D^{\mu_n} 
T^a_{IJ} \psi^J - \mbox{trace terms}
\end{equation}
where ${\cal S}$ denotes symmetrization on the Lorentz indices. Although this
has already been treated in $1/\Nf$ in \cite{14} its value forms part of the 
flavour singlet calculation detailed later. The full critical exponent 
associated with (3.1) is $\eta_{\footnotesize{\mbox{ns}}}^{(n)}$,  
\begin{equation} 
\eta_{\footnotesize{\mbox{ns}}}^{(n)} ~=~ \eta ~+~ \eta_{\cal O} 
\end{equation} 
The first piece corresponds in exponent language to the wave function 
renormalization of the constituent fields of (3.1). The second part reflects 
the renormalization of the operator itself. Although each term of (3.2) is 
gauge dependent the combination is gauge independent. In perturbation theory 
the renormalization is carried out by inserting (3.1) in some Green's function
and examining its divergence structure. Here we determine the scaling 
behaviour of the integrals where ${\cal O}_{\mbox{\footnotesize{ns}}}$ is 
inserted in a quark $2$-point function. The two leading order Feynman diagrams
are given in fig 1. With the regularization each graph is evaluated with the
critical propagators (2.5) in $d$-dimensions. As in perturbative calculations
\cite{3} we project the Lorentz indices of the operator into a basis using a 
null vector $\Delta_\mu$, with $\Delta^2$ $=$ $0$. (This is not to be confused
with the regularizing parameter $\Delta$ which is a scalar object.) They have
the general form, omitting the external momentum dependence,  
\begin{equation} 
\frac{X}{\Delta} ~+~ Y ~+~ O(\Delta) 
\end{equation} 
where $X$ and $Y$ are functions of $d$. The integrals are straightforward to
compute using standard rules for massless integrals. To obtain the leading 
order large $\Nf$ contribution $\alpha$ and $\beta$ are replaced by $\mu$ and 
$1$ respectively. Following \cite{19} the residue $X$ of each graph contributes
to $\eta_{{\cal O},1}^{(n)}$. In this instance we have for the respective 
graphs   
\[ 
\frac{2\mu C_2(R)\eta^{\mbox{o}}_1}{(\mu-2)(2\mu-1)T(R)} \left[ 1 ~-~ b ~-~ 
\frac{(\mu-1)^3}{(\mu+n-1)(\mu+n-2)} \right]  
\] 
and 
\begin{equation}
\frac{4\mu(\mu-1)C_2(R)\eta^{\mbox{o}}_1}{(\mu-2)(2\mu-1)T(R)}
\sum_{l=2}^n \frac{1}{(\mu+l-2)}
\end{equation}
where we have included a factor of $2$ in the second to account for the 
contibution of the mirror image and used the value of $z$, (2.14). Summing the
contributions yields, \cite{14},  
\begin{equation}
\eta^{(n)}_{{\footnotesize{\mbox{ns}}},1} ~=~ 
\frac{2C_2(R)(\mu-1)^2\eta^{\mbox{o}}_1}{(2\mu-1)(\mu-2)T(R)} 
\left[ \frac{(n-1)(2\mu+n-2)}{(\mu+n-1)(\mu+n-2)} ~+~ \frac{2\mu}{(\mu-1)} 
[\psi(\mu+n-1) \, - \, \psi(\mu)] \right] 
\end{equation}
where $\psi(x)$ is the logarithmic derivative of the $\Gamma$-function. We 
recall that this result is in agreement with all known perturbative $\MSbar$ 
results to three loops, \cite{3,7}. In concentrating on the detail for this 
operator we will follow the same procedure in the remainder of the paper with
minimal comment. 

We now turn to the treatment of the flavour singlet twist-$2$ operators. Before
analysing at the fixed point we need to recall several features of their 
perturbative renormalization. First, the operators are, \cite{25,4}, 
\begin{eqnarray}
{\cal O}^{\mu_1 \ldots \mu_n}_{\mbox{\footnotesize{F}}} &=&  
i^{n-1} {\cal S} \bar{\psi}^I \gamma^{\mu_1} D^{\mu_2} \ldots D^{\mu_n} 
\psi^J - \mbox{trace terms} \\ 
{\cal O}^{\mu_1 \ldots \mu_n}_{\mbox{\footnotesize{G}}} &=&  
\half i^{n-2} {\cal S} \, \mbox{tr} \, G^{a \, \mu_1\nu} D^{\mu_2} 
\ldots D^{\mu_{n-1}} G^{a \, ~ \mu_n}_{~~\nu} - \mbox{trace terms}
\end{eqnarray}
As each operator has the same dimension in four dimensions and quantum numbers
they mix under renormalization. In other words defining the vector ${\cal O}_i$
$=$ $\{ {\cal O}_F, {\cal O}_G \}$ then the bare and renormalized operators are
related by 
\begin{equation} 
{\cal O}^{\footnotesize{\mbox{ren}}}_i ~=~ Z_{ij} 
{\cal O}^{\footnotesize{\mbox{bare}}}_j 
\end{equation} 
where $Z_{ij}$ is a $2$ $\times$ $2$ matrix of renormalization constants. 
Consequently the associated anomalous dimension is a $2$ $\times$ $2$ matrix
$\gamma_{ij}(g)$. It has the following structure, with the $\Nf$ dependence 
explicit, 
\begin{equation} 
\gamma_{ij}(g) ~=~ \left( 
\begin{array}{ll} 
\gamma^{qq} & \gamma^{gq} \\ 
\gamma^{qg} & \gamma^{gg} \\ 
\end{array} 
\right) 
{}~=~ \left( 
\begin{array}{ll} 
a_1g + (a_2\Nf + a_3)g^2 & b_1g + (b_2\Nf + b_3)g^2 \\  
c_1\Nf g + c_2\Nf g^2 & (d_1\Nf + d_2)g + (d_3\Nf + d_4)g^2 \\  
\end{array} 
\right) 
\end{equation} 
where, for example, 
\begin{eqnarray} 
a_1 &=& 2C_2(R) \left[ 4 S_1(n) ~-~ 3 ~-~ \frac{2}{n(n+1)} \right] ~~~,~~~  
b_1 ~=~ -~ \frac{4(n^2+n+2)C_2(R)}{n(n^2-1)} \nonumber \\
c_1 &=& -~ \frac{8(n^2+n+2)T(R)}{n(n+1)(n+2)} ~~~,~~~  
d_1 ~=~ \frac{8}{3} T(R) \nonumber \\ 
a_2 &=& T(R)C_2(R) \left[ \frac{4}{3} ~-~ \frac{160}{9}S_1(n) ~+~ 
\frac{32}{3}S_2(n) \right. \nonumber \\ 
&&+~ \left. \frac{16[11n^7+49n^6+5n^5-329n^4-514n^3-350n^2-240n-72]} 
{9n^3(n+1)^3(n+2)^2(n-1)} \right] \nonumber \\ 
b_2 &=& \frac{32C_2(R)T(R)}{3} \left[ \frac{1}{(n+1)^2} ~+~ \frac{(n^2+n+2)}
{n(n^2-1)} \left( S_1(n) ~-~ \frac{8}{3} \right) \right] 
\end{eqnarray} 
The remaining entries of (3.9) can be found in, for instance, \cite{4,12} and 
are not important for the present situation. To compute $\gamma_{ij}(g)$ the
operators are inserted in both quark and gluon $2$-point functions. Various
one loop graphs which occur are illustrated in figs 1 and 2. Clearly from (3.9)
the $\Nf$ dependence is not the same in each term. For example, at $g_c$ the
$\Nf$ dependence of each entry is respectively, $O(1/\Nf)$, $O(1/\Nf)$, $O(1)$
and $O(1)$. Further, in practical applications it is sometimes useful to 
compute with the operator eigenbasis of (3.9) which simplifies the RGE 
involving $\gamma_{ij}(g)$ and therefore the evolution of the Wilson 
coefficients of the operator product expansion. From (3.9) this leads to the 
eigenvalues  
\begin{eqnarray} 
\lambda_\pm &=& \frac{1}{2} ( d_1\Nf ~+~ a_1 ~+~ d_2 ~ \pm ~ \sqrt{A_1}) g 
\nonumber \\ 
&&+~ \frac{1}{2} \left( (a_2+d_3)\Nf ~+~ a_3 ~+~ d_4 ~ \pm ~ 
\frac{A_2}{2\sqrt{A_1}} \right) g^2 ~+~ O(g^3) 
\end{eqnarray} 
where 
\begin{eqnarray} 
A_1 &=& d_1^2 \Nf^2\left[ 1 ~+~ \frac{2(d_4-a_1)}{d_1\Nf} ~+~ \frac{4b_1c_1}
{d^2_1\Nf^2} \right] \nonumber \\ 
A_2 &=& 2\Nf [ (d_1(d_3-a_2) + 2c_1b_2)\Nf \\  
&&+~ (d_2-a_1)(d_3-a_2) + d_1(d_4-a_3) + 2(c_1b_3+c_2b_1) ] \nonumber  
\end{eqnarray}  
Or evaluating at $g_c$ the related eigenexponents are at leading order in large
$\Nf$ 
\begin{eqnarray} 
\lambda_+ &=& d_1 \Nf g \nonumber \\ 
\lambda_- &=& \left( a_1 - \frac{b_1c_1}{d_1}\right) g ~+~ 
\left( a_2 - \frac{b_2c_1}{d_1}\right) g^2\Nf ~+~ O(\Nf^2 g^3) 
\end{eqnarray} 
Clearly the $\Nf$ dependence in each eigenexponent differs. The eigenoperators
associated with each eigenvalue, $\lambda_\pm$, are a combination of the 
original operators. For example, that associated with $\lambda_-$ has 
predominant contributions from the fermionic operator (3.6). Likewise 
$\lambda_+$ is associated primarily with (3.7). 

For the critical point analysis there will be a $2$ $\times$ $2$ matrix of 
critical exponents analogous to (3.9) which are computed by inserting the 
critical propagators into the diagrams of figs 1 and 2. In addition the graphs 
of fig 3 are also of the same order in $1/\Nf$. However in determining the 
contribution to $X$ of each of the graphs it turns out that several are trivial
due to the imbalance of the $\Nf$ dependence already mentioned. For instance 
the leading order term for $\lambda_+$ arises purely from the tree graph of fig
2. Therefore we take as its entry in $\eta_{ij}$ $\equiv$ $\gamma_{ij}(g_c)$ as 
\cite{24},  
\begin{equation} 
\eta_{\mbox{\footnotesize{GG}},1} ~=~ 2 \epsilon 
\end{equation} 
Also $\eta_{\mbox{\footnotesize{FG}},1}$ does not need to be evaluated as its 
leading order value is given purely by the one loop perturbative result. Next 
the contribution from the final graph of fig 2 is identically zero. That is, 
with (2.5) the graph is $\Delta$-finite. Therefore the only non-trivial entry 
to compute is $\eta_{\mbox{\footnotesize{FF}},1}$. As the non-singlet part has 
already been determined this reduces to evaluating the two loop graphs of fig 
3. Each is $b$-independent and respectively contribute, for even $n$,  
\begin{eqnarray}  
&&-~ \frac{\mu(\mu-1)\Gamma(n)\Gamma(2\mu)\eta^{\mbox{o}}_1} 
{(\mu-2)(2\mu-1)(\mu+n-1)(\mu+n-2)\Gamma(2\mu-1+n)} \nonumber \\ 
&&~~~ \times [n(n(n-2) + 2(\mu-2+n)(2\mu-3)+(2\mu-2+n)) + 2(n-2)(\mu+n-1)]
\nonumber 
\end{eqnarray}
and  
\begin{equation} 
\frac{8\mu(\mu-1)\Gamma(n-1)\Gamma(2\mu)C_2(R)\eta^{\mbox{o}}_1}
{(\mu-2)(2\mu-1)\Gamma(2\mu-1+n)T(R)} 
\end{equation} 
Hence,  
\begin{eqnarray}
\eta_{{\footnotesize{\mbox{FF}}},1}^{(n)} 
&=& \frac{(\mu-1)C_2(R)\eta^{\mbox{o}}_1}{(2\mu-1)
(\mu-2)T(R)\Nf} \left[ \frac{2(\mu-1)(n-1)(2\mu+n-2)}{(\mu+n-1)(\mu+n-2)} 
{}~+~ 4\mu[\psi(\mu-1+n) - \psi(\mu)] \right. \nonumber \\
&&-~ \left. \frac{\mu\Gamma(n-1)\Gamma(2\mu)}{(\mu+n-1)(\mu+n-2)
\Gamma(2\mu-1+n)} \right. \nonumber \\ 
&&~~~~ \times \left. [(n^2+n+2\mu-2)^2 + 2(\mu-2)(n(n-1)(2\mu-3+2n) 
+ 2(\mu-1+n))] \frac{}{} \right]
\end{eqnarray}

We now discuss the structure of $\eta_{ij}$. Unlike the perturbative mixing 
matrix $\gamma_{ij}(g)$, $\eta_{ij}$ is triangular. At first sight this would 
appear to be inconsistent with perturbation theory. However, at leading order 
in $1/\Nf$ the calculation of $\eta_{ij}$ in fact determines the critical 
anomalous dimensions of the eigenoperators {\em directly}. This is not 
unexpected if one studies the dimensions of (3.6) and (3.7) at $g_c$. There 
clearly the canonical dimensions of each operator is different and therefore 
there is no mixing. The vanishing of certain graphs of fig 2 is merely a 
reflection of this in the large $\Nf$ calculation. This indirect relation 
between the exponents of the eigenoperators (3.6) and (3.7) is the reason why 
we distinguish the perturbative entries of (3.9) by $q$ and $g$ in contrast to
$F$ and $G$ for the eigenoperators. A further justification of this point of 
view comes from the comparison of the coefficients of the $O(\epsilon)$ and 
$O(\epsilon^2)$ terms in the expansion of (3.16) with $\lambda_-$ evaluated to 
the same order at $g_c$. We have checked that they are in total agreement with 
(3.10) for all $n$. A further check is that the anomalous dimension must vanish
at $n$ $=$ $2$. Then the original operator corresponds to a conserved physical 
quantity, the energy momentum tensor which has zero anomalous dimension. From 
(3.16) it is easy to check that $\eta_{{\footnotesize{\mbox{FF}}},1}^{(2)}$ $=$
$0$. 

It is worth commenting on this calculation in relation to the NATM and QCD 
equivalence noted earlier, \cite{22}. Clearly $\lambda_-$ and 
$\eta_{{\footnotesize{\mbox{FF}}}}$ contain contributions from the insertion of
gluonic operators in a Green's function. However the graphs we evaluate to 
obtain $\eta_{{\footnotesize{\mbox{FF}}}}$ involve only (3.6). The resolution 
of this apparent inconsistency is obtained by studying the integration of each 
quark loop in fig 3 with (2.5) and the $\epsilon$ expansion of the individual 
graphs. Clearly in perturbation theory the graphs of fig 1 will contribute to
the one loop renormalization whilst those of fig 3 will give part of the two 
loop value of the anomalous dimension. So one would expect the large $\Nf$ 
graphs to be $O(\epsilon)$ and $O(\epsilon^2)$ respectively. This is not the 
case. Studying (3.15) each graph of fig 3 is $O(\epsilon)$ and from (3.16) 
their sum is also of this order. The point is that after performing the quark 
loop integral and examining the resulting one loop integral, it contains a part 
which would correspond to the ordinary perturbation theory two loop value as 
well as a piece that corresponds to the final graph of fig 2 which is a 
{\em one} loop integral. In other words an effective gluonic operator like 
(3.7) emerges naturally in the exponent calculation. In effect we are 
confirming in our calculation the equivalence observed in \cite{22} where we 
recall that the three and four point gluon interactions were similarly 
reproduced by integrating out quark loops.  

We conclude this section by giving an indication of the $n$-dependence of at 
least the leading order $1/\Nf$ coefficients of higher loop terms in the series
for $\gamma_-(g)$. Having established the correctness of our expansion at two 
loops the higher order coefficients are  
\begin{eqnarray}
a_3 ~-~ \frac{b_3c_1}{d_1} &=& \frac{2}{9}S_3(n) ~-~ \frac{10}{27}S_2(n) ~-~
\frac{2}{27}S_1(n) ~+~ \frac{17}{72} ~-~ 
\frac{2(n^2+n+2)^2[S_2(n)+S^2_1(n)]}{3n^2(n+2)(n+1)^2(n-1)} \nonumber \\
&&-~ \frac{2S_1(n)(16n^7+74n^6+181n^5+266n^4+269n^3+230n^2+44n-24)} 
{9(n+2)^2(n+1)^3(n-1)n^3}  \nonumber \\
&&-~ [100n^{10}+682n^9+2079n^8+3377n^7+3389n^6+3545n^5+3130n^4 
\nonumber \\ 
&&~~~~~ + \, 118n^3-940n^2-72n+144]/[27(n+2)^3(n+1)^4n^4(n-1)]  
\end{eqnarray}
and 
\begin{eqnarray}
a_4 ~-~ \frac{b_4c_1}{d_1} &=& 
\frac{2}{27}S_4(n) ~-~ \frac{10}{81}S_3(n) ~-~ \frac{2}{81}S_2(n) ~-~ 
\frac{2}{81}S_1(n) ~+~ \frac{131}{1296} \nonumber \\ 
&&+~ \zeta(3) \left[ \frac{4}{27}S_1(n) - \frac{2}{27n(n+1)} - \frac{1}{9} 
- \frac{2(n^2+n+2)^2}{9n^2(n+2)(n+1)^2(n-1)} \right] \nonumber \\ 
&&-~ \frac{4(n^2+n+2)^2[2S_3(n) + 3S_2(n)S_1(n) + S_1^3(n)]} 
{27(n+2)n^2(n-1)(n+1)^2} \nonumber \\ 
&&+\, 2[S_2(n) + S_1^2(n)] \frac{(16n^7 \! + 74n^6 + 181n^5 + 266n^4
                 + 269n^3 + 230n^2 \! + 44n - 24)}{27n^3(n+2)^2(n+1)^3(n-1)} 
\nonumber \\  
&&-~ 2S_1(n)[88n^{10} + 608n^9 + 1947n^8 + 3405n^7 + 3670n^6 + 3693n^5 
\nonumber \\ 
&&~~~~~~~~~~~~~ + 2973n^4 - 8n^3 - 920n^2 - 48n + 144] 
/[81(n+2)^3(n+1)^4n^4(n-1)] \nonumber \\  
&&+~ [68n^{13} + 548n^{12} + 1861n^{11} + 2474n^{10} - 817n^9 - 4143n^8 
- 1712n^7 \nonumber \\ 
&&~~~~ - 2871n^6 - 7702n^5 - 2586n^4 + 3136n^3 + 1952n^2 \nonumber \\
&&~~~~ - 288n - 288]/[81(n+2)^4(n+1)^5(n-1)n^5]  
\end{eqnarray}
For future reference we list the values of (3.17) calculated for low moments in
table 1. A similar table was produced for the analogous coefficient in the 
non-singlet case. It is important to note that all the fractions up to $n$ $=$
$8$ are in {\em exact} agreement with the recent explicit three loop singlet 
results of \cite{7}, when allowance is made for different coupling constant 
definitions. 

\sect{$\gamma^5$.} 

To apply the large $\Nf$ method to polarized operators we need to review the
treatment of $\gamma^5$ in perturbation theory and earlier $1/\Nf$ 
calculations. As is well known one must be careful in arbitrary spacetime 
dimensions when $\gamma^5$ or the pseudotensor $\epsilon_{\mu\nu\sigma\rho}$ 
are present, \cite{26}. The simple reason is that both are purely four 
dimensional objects unlike, say, $\gamma^\mu$ and $\eta^{\mu\nu}$ and do not 
generalize in the arbitrary dimensional case. Therefore problems will arise in 
perturbation theory when one uses dimensional regularization. With this 
regularization calculations are performed in $d$ $=$ $4$ $-$ $2\epsilon$ 
dimensions where the infinities are removed before taking the $\epsilon$ 
$\rightarrow$ $0$ limit. There are, however, various ways of incorporating 
$\gamma^5$ in such calculations, [26-28]. (A review is, for example, 
\cite{29}.) The original approach of \cite{26} was to split the $d$-dimensional
spacetime into physical and unphysical complements. In the former subspace 
Lorentz indices run from $1$ to $4$ whilst they range over the remaining 
dimensions in the latter. So, for example, the $\gamma$-matrices are split into
two components 
\begin{equation} 
\gamma^\mu ~=~ \bar{\gamma}^\mu ~+~ \hat{\gamma}^\mu 
\end{equation}  
where the bar, $\bar{}~$, denotes the physical four dimensional spacetime and 
the hat, $\hat{}~$, the remaining $(d-4)$-dimensional subspace. Then the 
Clifford algebra reduces to 
\begin{equation} 
\{ \bar{\gamma}^\mu , \bar{\gamma}^\nu \} ~=~ 2 \bar{\eta}^{\mu\nu} ~~,~~ 
\{ \bar{\gamma}^\mu , \hat{\gamma}^\nu \} ~=~ 0 ~~,~~ 
\{ \hat{\gamma}^\mu , \hat{\gamma}^\nu \} ~=~ 2 \hat{\eta}^{\mu\nu}  
\end{equation} 
The anti-commutativity of $\gamma^5$ is not preserved in the full spacetime. 
Instead the following relations are used 
\begin{equation} 
\{ \bar{\gamma}^\mu , \gamma^5 \} ~=~0 ~~,~~ 
[ \hat{\gamma}^\mu , \gamma^5 ] ~=~ 0  
\end{equation} 
It is known that these definitions give a consistent method for treating 
$\gamma^5$, \cite{28}. Traces involving an odd number of $\gamma^5$'s are 
performed via, in our conventions, 
\begin{equation} 
\mbox{tr} ( \gamma^5 \gamma^\mu \gamma^\nu \gamma^\sigma \gamma^\rho ) ~=~ 
4 \bar{\epsilon}^{\mu\nu\sigma\rho} 
\end{equation} 
which acts like a projection into the physical dimensions. Further if (4.4) 
occurs in a loop integral where the $\gamma$-matrices are contracted with loop
momenta the integral is performed first and then the Lorentz index contractions
carried out, with the caveat that external momenta are physical, $\hat{p}_\mu$ 
$=$ $0$. 

For high order perturbative calculations this splitting of the algebra is not
always practical, \cite{30}. It would be easier if a $d$-dimensional 
calculation could be performed. Such an approach has been introduced in 
\cite{30,31} and carried out successfully for $3$-loop calculations in 
\cite{30}. The first step there is to replace $\gamma^5$ by 
\begin{equation} 
\gamma^5 ~=~ \frac{1}{4!} \epsilon_{\mu\nu\sigma\rho} \gamma^\mu \gamma^\nu 
\gamma^\sigma \gamma^\rho 
\end{equation} 
and remove the $\epsilon$-tensor from the renormalization procedure. The 
$\gamma$-matrices of (4.5) are treated as $d$-dimensional in the calculation
before projecting to the physical dimension. If two such $\epsilon$-tensors
are present then they can be replaced by a sum of products of $\eta$-tensors
which is treated as $d$-dimensional. One performs the renormalization in a 
minimal way as usual to determine the renormalization constants. To complete
the calculation, in relation to the $\MSbar$ scheme, one must introduce a 
finite renormalization constant $Z_5$ in addition to the first renormalization
constant in order to restore the Ward identity, \cite{31}. 

For the treatment of $\gamma^5$ in the $1/\Nf$ expansion we recall the simple 
example of the flavour non-singlet axial current. In \cite{33} the method 
outlined above was followed to correctly determine the anomalous dimension of 
${\cal O}^{\mu 5}_{\mbox{\footnotesize{ns}}}$ $=$ $\bar{\psi} \gamma^\mu  
\gamma^5 \psi$ at $O(1/\Nf)$. First, if one wishes to find the critical 
exponent associated with the non-singlet vector current 
${\cal O}^\mu_{\mbox{\footnotesize{ns}}}$ $=$ $\bar{\psi} \gamma^\mu \psi$ 
then it is inserted in a $2$-point function and the residue with respect to
$\Delta$ is determined. The only relevant graph at $O(1/\Nf)$ is the first 
graph of fig 1. If we insert the more general non-singlet operator $\bar{\psi} 
\Gamma \psi$ then the contribution to the critical exponent from the graph is 
\begin{equation} 
-~ \frac{ [ \gamma^\nu\gamma^\sigma \Gamma \gamma_\sigma \gamma_\nu ~-~ 
2\mu(1-b) \Gamma ] \eta^{\mbox{o}}_1}{2(2\mu-1)(\mu-2)T(R)} 
\end{equation} 
where the square brackets are understood to mean the coefficient of the matrix 
$\Gamma$ after all $\gamma$-matrix manipulations have been performed for an 
explicit form of $\Gamma$. Therefore for $\Gamma$ $=$ $\gamma^\mu$, (4.6) 
gives 
\begin{equation} 
-~ \frac{[(2\mu-1)(\mu-2) + b \mu] \eta^{\mbox{o}}_1 }{(2\mu-1)(\mu-2)} 
\end{equation} 
and so with (2.12) and (2.13)  
\begin{equation} 
\eta_{{\cal O}^\mu_{\mbox{\footnotesize{ns}}}} ~=~ 0 
\end{equation} 
consistent with the Ward identity in exponent language, \cite{20}. For 
${\cal O}^{\mu 5}_{\mbox{\footnotesize{ns}}}$ one performs the $\gamma$-algebra
of (4.6) using (4.2), to give 
\begin{equation} 
-~ \frac{[(2\mu-9)(\mu-2) + b\mu]\eta^{\mbox{o}}_1}{(2\mu-1)(\mu-2)T(R)} 
\end{equation} 
Thus 
\begin{equation} 
\tilde{\eta}_{{\cal O}^{\mu 5}_{\mbox{\footnotesize{ns}}}} ~=~  
\frac{8\eta^{\mbox{o}}_1}{(2\mu-1)T(R)} 
\end{equation} 
where $\tilde{}$ denotes that the object still has to be augmented by the 
finite renormalization. As discussed in \cite{33} this does not preserve four 
dimensional chiral symmetry and is not consistent with the Ward identity. To 
proceed correctly we need to include a finite renormalization constant. In 
\cite{33} this was computed to be 
\begin{equation} 
Z_5 ~=~ 1 ~+~ \frac{C_2(R)\epsilon}{6T(R)\Nf} \hat{\mbox{L}} \left\{ 
\frac{\ln [1 - 4T(R)\Nf a_{\mbox{\footnotesize{S}}}/(3\epsilon)]} 
{B(2-\epsilon,2-\epsilon) B(3-\epsilon,1+\epsilon)} \right\} ~+~ 
O \left( \frac{1}{\Nf^2} \right) 
\end{equation} 
where $\hat{\mbox{L}}$ is the Laurent operator which removes non-singular terms
from the expansion of the braces and $B(x,y)$ is the Euler $\beta$-function. 
The constant $Z^{\mbox{\footnotesize{ns}}}_5$ is defined from the requirement 
that, \cite{30},  
\begin{equation} 
Z^{\mbox{\footnotesize{ns}}}_5 ~ {\cal R}_{\footnotesize{\MSbar}} ~ \langle 
\bar{\psi} \, {\cal O}^{\mu 5}_{\mbox{\footnotesize{ns}}} \, \psi \rangle ~=~ 
\gamma^5 \, {\cal R}_{\footnotesize{\MSbar}} ~ \langle \bar{\psi} \, 
{\cal O}^\mu_{\mbox{\footnotesize{ns}}} \, \psi \rangle  
\end{equation} 
where ${\cal R}_{\footnotesize{\MSbar}}$ denotes the $R$-operator or 
renormalization procedure. In other words the anti-commutativity of $\gamma^5$
is restored by this condition. Using the information in this finite 
renormalization together with (4.10) the correct $\MSbar$ anomalous dimension
does emerge to all orders in the coupling at $O(1/\Nf)$.  

There are several disadvantages, however, with the form of (4.11). First, it is
not as compact as the $O(1/\Nf)$ exponents that have been produced in earlier 
work, \cite{16}. Second by examining (4.11) the result can be simplified since 
the construction of $Z^{\mbox{\footnotesize{ns}}}_5$ is in effect equivalent to
the difference of the exponents (4.7) and (4.9) at $O(1/\Nf)$. In other words 
the contribution from the finite renormalization to the final $\MSbar$ exponent
is equal to  
\begin{equation} 
-~ \frac{8\eta^{\mbox{o}}_1}{(2\mu-1)T(R)} 
\end{equation} 
Thus the sum of (4.11) and (4.13) correctly gives in $\MSbar$  
\begin{equation} 
\eta_{{\cal O}^{\mu 5}_{\mbox{\footnotesize{ns}}}} ~=~ 0 
\end{equation} 
Another difficulty with this procedure is that there is a quicker derivation 
based on features of the fixed point approach. In perturbation theory the 
regularization used is dimensional in contrast to the critical point method. 
There the spacetime dimension is fixed and the regularization is analytic as it
is the gluon dimension which is adjusted. The upshot is that, at least for 
non-singlet currects, one can use the anti-commutativity of $\gamma^5$ in 
$d$-dimensions. Therefore with $\Gamma$ $=$ $\gamma^\mu\gamma^5$ in (4.6) 
anti-commuting $\gamma^5$ twice immediately gives the same contribution as 
$\Gamma$ $=$ $\gamma^\mu$. Hence the $\MSbar$ result (4.14) follows directly. 
We have checked this procedure explicitly for other non-singlet operators such 
as $\bar{\psi} \gamma^5 \psi$ and ${\cal S} \bar{\psi} \gamma^5 \gamma^{\mu_1} 
D^{\mu_2} \ldots D^{\mu_n} \psi$ by calculating the analogous finite 
renormalization constant from a condition similar to (4.12) and observing that 
the result agrees with the direct anti-commuting $\gamma^5$ calculation. So, 
for example, the unpolarized and polarized non-singlet twist-$2$ operators have
the same anomalous dimensions, (3.5). In other words we have justified the use 
of an anti-commuting $\gamma^5$ in non-singlet sectors of calculations. 
Although much of the content of this section may appear straightforward, there 
is an important lesson in the result (4.11) from \cite{33} for singlet 
operators. Then closed quark loops with an odd number of $\gamma^5$ matices 
will occur which means quantities like (4.11) will need to be computed. As we 
have demonstrated that this is equivalent to the difference in the anomalous 
dimensions of the operators of the renormalization condition (4.12) defining 
the finite renormalization constant, flavour singlet operators can be handled 
in an efficient way. We will come back to this point in a later section.  

\sect{Polarized singlet operators.} 

We now extend the unpolarized singlet calculation of section 3 to the polarized
case as it is important to compare with recent perturbative calculations 
[9-11]. The twist-$2$ operators are \cite{8}, 
\begin{eqnarray} 
{\cal O}_F^{\mbox{\footnotesize{pol}}} &=& i^{n-1} {\cal S} \bar{\psi} \gamma^5 
\gamma^{\mu_1} D^{\mu_2} \ldots D^{\mu_n} \psi - \mbox{trace terms} \\  
{\cal O}_G^{\mbox{\footnotesize{pol}}} &=& \half i^{n-2} {\cal S} 
\epsilon^{\mu_1\alpha\beta\gamma} \, \mbox{tr} \, G_{\beta\gamma} D^{\mu_2} 
\ldots D^{\mu_{n-1}} G^{\mu_n}_{~~\, \alpha} - \mbox{trace terms}
\end{eqnarray} 
Several features of the computation of the critical exponents will parallel
section 3 such as the triangularity of the mixing matrix and the $\Nf$ 
dependence of $\gamma^{\mbox{\footnotesize{pol}}}_{ij}(g)$. The essential 
difference is the effect $\gamma^5$ has in the two two loop graphs of fig 3 
which we focus on here. The contribution from the graphs of fig 1 is the same
as (3.15). 

With (4.4) the second graph of fig 3 is $\Delta$-finite and gives no 
contribution to 
$\eta^{\mbox{\footnotesize{pol}}}_{{\mbox{\footnotesize{FF}}},1}$. For the 
other graph one can compute the quark loop in $d$-dimensions before carrying
out the second loop integral, also in arbitrary dimensions. The projection to 
four dimensions is made at the end. Adding all pieces we have,  
\begin{eqnarray} 
\eta^{\mbox{\footnotesize{pol}}}_{{\mbox{\footnotesize{FF}}},1} &=& 
\frac{2C_2(R)\eta^{\mbox{o}}_1}{(2\mu-1)(\mu-2)T(R)} \left[  
\frac{(n-1)(2\mu+n-1)(\mu-1)^2}{(\mu+n-1)(\mu+n-2)} \right. \\ 
&&\left. +~ 2\mu(\mu-1) [\psi(\mu-1+n) - \psi(\mu)] ~-~ 
\frac{\mu(2\mu+n-5)(n+2)\Gamma(n)\Gamma(2\mu)} 
{2(\mu+n-1)(\mu+n-2)\Gamma(2\mu+n-2)} \right] \nonumber  
\end{eqnarray} 
As in section 3, due to the $\Nf$ dependence we have 
\begin{equation} 
\eta^{\mbox{\footnotesize{pol}}}_{{\mbox{\footnotesize{GG}}},1} ~=~ 
2\epsilon 
\end{equation}  
We have checked that the $\epsilon$-expansion of (5.3) agrees with the 
anomalous dimension of the predominantly fermionic eigenoperator of the mixing
matrix at two loops, [9-11]. For completeness we note in the notation of (3.9), 
\begin{eqnarray} 
a_1^{\mbox{\footnotesize{pol}}} &=& 2C_2(R) \left[ 4 S_1(n) ~-~  
3 ~-~ \frac{2}{n(n+1)} \right] ~~~,~~~  
b_1^{\mbox{\footnotesize{pol}}} ~=~ -~ \frac{4(n+2)C_2(R)}{n(n+1)} 
\nonumber \\
c_1^{\mbox{\footnotesize{pol}}} &=& -~ \frac{8(n-1)T(R)}{n(n+1)} ~~~,~~~  
d_1^{\mbox{\footnotesize{pol}}} ~=~ \frac{8}{3} T(R) \nonumber \\ 
a_2^{\mbox{\footnotesize{pol}}} &=& T(R)C_2(R) \left[ \frac{4}{3} ~-~ 
\frac{160}{9}S_1(n) ~+~ \frac{32}{3}S_2(n) ~+~ 
\frac{32[10n^4+17n^3+10n^2+21n+9]}{9n^3(n+1)^3} \right] \nonumber \\ 
b_2^{\mbox{\footnotesize{pol}}} &=& - \, \frac{32(n+2)C_2(R)T(R)}{3n(n+1)} 
\left[ S_1(n) ~-~ \frac{8}{3} ~+~ \frac{(n+2)}{(n+1)} \right] 
\end{eqnarray} 
This agreement, moreover, justifies our treatment of $\gamma^5$ at the fixed
point to be an anti-commuting object whose appearance in closed loops is 
treated with (4.4). Finally we deduce  
\begin{eqnarray}
\left[ a_3 ~-~ \frac{b_3c_1}{d_1} \right]^{\mbox{\footnotesize{pol}}} 
&=& \frac{2}{9}S_3(n) ~-~ \frac{10}{27}S_2(n) ~-~ \frac{2}{27}S_1(n) \,+\,  
\frac{17}{72} - \frac{2(n+2)(n-1)[S_2(n)+S^2_1(n)]}{3n^2(n+1)^2} \nonumber \\
&&+~ \frac{2S_1(n)(13n^3-6n^2+2n+3)(n+2)}{9(n+1)^3n^3} \\
&&-~ \frac{[61n^6+83n^5+27n^4+217n^3+68n^2-36n-18]}{27(n+1)^4n^4} \nonumber  
\end{eqnarray}
and 
\begin{eqnarray}
\left[ a_4 ~-~ \frac{b_4c_1}{d_1} \right]^{\mbox{\footnotesize{pol}}} 
&=& \frac{2}{27}S_4(n) ~-~ \frac{10}{81}S_3(n) ~-~ \frac{2}{81}S_2(n) ~-~ 
\frac{2}{81}S_1(n) ~+~ \frac{131}{1296} \nonumber \\ 
&&+~ \zeta(3) \left[ \frac{4}{27}S_1(n) - \frac{2}{27n(n+1)} - \frac{1}{9} 
- \frac{2(n+2)(n-1)}{9n^2(n+1)^2} \right] \nonumber \\ 
&&-~ \frac{4(n+2)(n-1)[2S_3(n) + 3S_2(n)S_1(n) + S_1^3(n)]}{27n^2(n+1)^2} 
\nonumber \\  
&&+~ \frac{2(13n^3 - 6n^2 + 2n + 3)(n+2)[S_2(n) + S_1^2(n)]}{27n^3(n+1)^3} \\  
&&-~ \frac{2S_1(n)(49n^5 - 29n^4 + 95n^3 + 41n^2 - 15n - 9)(n+2)} 
{81(n+1)^4n^4} \nonumber \\  
&&+~ \frac{(19n^8 - 46n^7 + 3n^6 + 195n^5 - 392n^4 - 317n^3 - 32n^2 + 54n 
+ 18)}{81(n+1)^5n^5} \nonumber  
\end{eqnarray}
In the second column of table 1 we have evaluated (5.6) for low moments as a 
check for future three loop calculations. Although we have given exact 
fractions for the coefficients at three loops, the numerical values of the 
polarized and unpolarized entries do not differ significantly as $n$ increases. 

\sect{Singlet axial current.} 

Having considered a variety of fermionic operators which are both flavour 
non-singlet and singlet we turn to the remaining current. The renormalization
of the singlet axial current ${\cal O}^{\mu 5}_{\mbox{\footnotesize{s}}}$ $=$ 
$\bar{\psi} \gamma^\mu \gamma^5 \psi$ is somewhat special. Unlike the singlet 
vector current the conservation of ${\cal O}^{\mu 5}_{\mbox{\footnotesize{s}}}$ 
is spoiled at the {\em quantum} level by the chiral anomaly, [34-36]. 
Consequently under renormalization the composite operator can develop a 
non-zero anomalous dimension. By contrast the conservation of the vector 
current ensures it has a zero anomalous dimension at all orders in the coupling
constant. Before attacking the problem of computing the $O(1/\Nf)$ exponent 
for ${\cal O}^{\mu 5}_{\mbox{\footnotesize{s}}}$ in the $\MSbar$ scheme, it is
worthwhile reviewing the perturbative approach \cite{32} and in particular 
\cite{30}. (Other related contributions to the renormalization of the axial 
anomaly are [37-39].) In the three loop analysis, \cite{30}, two 
renormalization constants are determined in a manner described earlier for 
other currents. One is the renormalization constant which removes the 
infinities in the usual way but using the standard $\gamma$-algebra and the
definition of $\gamma^5$, (4.5). This renormalization does not preserve the 
axial anomaly, in operator form, in four dimensions. To remedy this a second
finite renormalization constant $Z_5^{\mbox{\footnotesize{anom}}}$ is 
required. The relevant constraint in the present instance is determined by
ensuring that the operator form of the anomaly, [34-36],  
\begin{equation} 
\partial_\mu {\cal O}^{\mu 5}_{\mbox{\footnotesize{s}}} ~=~  
\frac{T(R)\Nf}{4g} ~ \epsilon^{\mu\nu\sigma\rho} G^a_{\mu\nu} G^a_{\sigma\rho} 
\end{equation} 
is preserved, leading to, \cite{36,29}, 
\begin{equation} 
Z_5^{\mbox{\footnotesize{anom}}} ~ {\cal R}_{\footnotesize{\MSbar}} ~  
\langle A \, \partial_\mu {\cal O}^{\mu 5}_{\mbox{\footnotesize{s}}} \, 
A \rangle ~=~ \frac{T(R)\Nf}{4g} \, {\cal R}_{\footnotesize{\MSbar}} ~  
\langle A \, \epsilon^{\mu\nu\sigma\rho} G^a_{\mu\nu} G^a_{\sigma\rho} \, 
A \rangle  
\end{equation} 

The large $\Nf$ calculation follows this two stage approach. In other words the
exponent corresponding to the $\MSbar$ anomalous dimension of 
${\cal O}^{\mu 5}_{\mbox{\footnotesize{s}}}$ is given by  
\begin{equation} 
\eta_{\mbox{\footnotesize{s}}} ~=~ \eta ~+~ \eta_{5,\mbox{\footnotesize{s}}} 
{}~+~ \eta_5^{\mbox{\footnotesize{fin}}} 
\end{equation} 
It is straightforward to compute the first graph of fig 3 in $d$-dimensions 
with the rules given previously. With (4.9) 
\begin{equation} 
\eta_1 ~+~ \eta_{5,\mbox{\footnotesize{s}},1} ~=~ 
\frac{C_2(R)\eta^{\mbox{o}}_1}{T(R)} \left[ \frac{8}{(2\mu-1)} ~-~ 
\frac{6}{(\mu-1)} \right] 
\end{equation}  
where the $b$-dependence has cancelled. We have used a split $\gamma$-algebra 
here to be consistent with the treatment of closed fermion loops in determining 
$\eta_5^{\mbox{\footnotesize{fin}}}$. There the first graphs of fig 1 and 3 
will occur as subgraphs. We have checked that the $\epsilon$-expansion of (6.4)
agrees with the three loop result for the same quantity in \cite{30}. 

To compute $\eta_5^{\mbox{\footnotesize{fin}}}$ we use the result of section 4.
There with a split $\gamma$-algebra the finite renormalization exponent was 
determined from the difference in the anomalous dimensions of the operators 
arising in the defining relation. In that case the restoration of the Ward 
identity was simple in that the result obtained was equivalent to using a fully
anti-commuting $\gamma^5$ initially and the operators themselves were similar 
in nature. For $\eta_5^{\mbox{\footnotesize{fin}}}$ the exponents of  
$\partial_\mu{\cal O}^{\mu 5}_{\mbox{\footnotesize{s}}}$ and $G$ $=$ 
$\epsilon^{\mu\nu\sigma\rho} G^a_{\mu\nu} G^a_{\sigma\rho}$, which are total 
derivatives must be determined separately in $d$-dimensions. In detailing that
calculation we focus on 
$\partial_\mu{\cal O}^{\mu 5}_{\mbox{\footnotesize{s}}}$ first.  

We insert $\partial_\mu{\cal O}^{\mu 5}_{\mbox{\footnotesize{s}}}$ into a gluon
$2$-point function as illustrated in fig 4. For the moment we take the 
momentum flow to be $p$ into the left gluon leg and $(p-q)$ out through the
other. This leaves a net flow of $q$ through the operator insertion which is
needed since a non-zero momentum must contract with $\gamma^\mu\gamma^5$ in
momentum space. To simplify the calculation of each integral we differentiate
with respect to $q_\phi$ and contract with $\epsilon_{\lambda\psi\theta\phi}
p^\theta$ where $\lambda$ and $\psi$ are the Lorentz indices of the gluon legs.
Then $q$ is set to zero, \cite{30}. This procedure ensures that part of the 
integrals contributing to the renormalization of the operator is projected out.
We have given the $O(1/\Nf)$ diagrams in figs 5 and 6. The former is the one
loop anomaly and with the critical propagators it is $\Delta$-finite. 
However, as the remaining graphs represent the higher order corrections the
value of the first graph of fig 5 must be factored off each to leave a formal 
sum of terms 
\begin{equation} 
-~ \frac{6T(R)\Nf (2\mu-1)(\mu-2)}{(\mu-1)} \left[ 1 ~+~ \frac{1}{\Nf} 
\left( \frac{X}{\Delta} ~+~ O(1) \right) \right] 
\end{equation} 
The overall factor is the $d$-dimensional value of the anomaly which is 
non-zero in four dimensions. In (6.5) we have included $z_1$ from the 
amplitudes of the quark fields which explains the origin of the factor 
$(\mu-2)$. The residue $X$ is the value of the $O(1/\Nf)$ part of the 
dimension of  $\partial_\mu{\cal O}^{\mu 5}_{\mbox{\footnotesize{s}}}$ we 
require. 

With the momentum flow as indicated we have computed the value of each graph
of fig 6. No graphs have been included where the vertex with the external 
gluon is dressed. These graphs together with the vertex counterterm do not
contribute to $X$ as they are $\Delta$-finite in sum. With the critical
propagators only the first two graphs are non-zero and give 
\begin{equation}
X ~=~ -~ \frac{C_2(R)\eta^{\mbox{o}}_1}{T(R)} \left[ 
\frac{[(2\mu-9)(\mu-2) + b\mu]}{(2\mu-1)(\mu-2)} ~+~ \frac{3}{(\mu-1)} \right] 
\end{equation} 
The remaining graphs are each $\Delta$-finite and we note the colour factors
of the last two graphs are each $C_2(G)$. Recalling the field content of  
$\partial_\mu{\cal O}^{\mu 5}_{\mbox{\footnotesize{s}}}$ we have  
\begin{equation}
\eta_{\partial {\cal O},1} ~=~ -~ \frac{C_2(R)\eta^{\mbox{o}}_1}{T(R)} \left[ 
\frac{8}{(2\mu-1)} ~-~ \frac{3}{(\mu-1)} \right] 
\end{equation} 

The treatment of $G$ is parallel to that just outlined. With the same 
projection of momenta the tree graph of fig 5 gives the normalization value of 
$(-$ $6)$ analogous to that of (6.5). The relevant graphs are given in fig 7 
and we list their respective contributions to $X$ as 
\begin{eqnarray} 
&-& \frac{C_2(R)\eta^{\mbox{o}}_1}{T(R)} ~~,~~ 
-~ \frac{[2C_2(R) - C_2(G)]\eta^{\mbox{o}}_1}{T(R)} ~~,~~ 
\frac{C_2(G)[4\mu^2-6\mu+1+b]\eta^{\mbox{o}}_1}{2(2\mu-1)(\mu-2)T(R)} 
\nonumber \\ 
&-& \frac{C_2(G) [8\mu^2-13\mu+4-\mu (1-b)] \eta^{\mbox{o}}_1 } 
{2(2\mu-1)(\mu-2)T(R)} 
\end{eqnarray} 
Useful in carrying out this calculation was the symbolic manipulation 
programme {\sc Form}, \cite{40}. The value of the three loop graph accounted
for the most tedious part of the calculation. However, we made use in part of
results of integrals which arose in the computation of the QCD 
$\beta$-function, \cite{41}. This was achieved by computing the dimension of
the composite operator $(G^a_{\mu\nu})^2$ associated with the coupling 
constant in a gluon $2$-point function. We have included a non-zero $b$ to 
observe its cancellation as a minor calculational check. Although the graphs
involved in computing the dimension of $G$ in fig 7 are similar in topology to 
those for $\partial {\cal O}^{\mu 5}_{\mbox{\footnotesize{s}}}$ in fig 6, the
values obtained are somewhat different. For example, the last graphs of each
figure are similar once the loop integral with the singlet current insertion
is performed which leaves a Feynman integral with an effective $G$ insertion.
The difference in the values arises due to the critical propagators used and
the fact that this loop integral changes the dimension of the gluon lines 
contracted with it and therefore the nature of the remaining loop integrations. 
One check on this is that the leading terms in the $\epsilon$ expansion of
each graph ought to agree. It is easy to observe that the first two values of
(6.8) give the same leading coefficient as the second term of (6.6). Likewise
the remaining two terms of (6.8) are $O(\epsilon)$. 

With the field content dimension (2.12) and (2.13), we find  
\begin{equation} 
\eta_{\mbox{\footnotesize{G}},1} ~=~ -~ \frac{3C_2(R)\eta^{\mbox{o}}_1}{T(R)} 
\end{equation} 
It is reassuring to note the cancellation of the terms involving $C_2(G)$ again
as the overall $\MSbar$ renormalization of  
${\cal O}^{\mu 5}_{\mbox{\footnotesize{s}}}$ at $O(1/\Nf)$ is expected to be
proportional to $C_2(R)$ only.  

With (6.7) and (6.9) the finite renormalization exponent is  
\begin{equation} 
\eta_{5,1}^{\mbox{\footnotesize{fin}}} ~=~ -~ \frac{C_2(R)\eta^{\mbox{o}}_1} 
{T(R)} \left[ \frac{8}{(2\mu-1)} ~+~ \frac{3(\mu-2)}{(\mu-1)} \right] 
\end{equation} 
Therefore 
\begin{equation} 
\eta_{\mbox{\footnotesize{s}},1} ~=~ -~ \frac{3\mu C_2(R)\eta^{\mbox{o}}_1} 
{(\mu-1)T(R)} 
\end{equation} 
where the cancellation of the terms proportional to $8/(2\mu-1)$ reflects the 
non-singlet calculation of section 4. A final check on this relatively simple
result is that it correctly reproduces the large $\Nf$ leading order two and
three loop $\MSbar$ coefficients of \cite{30,32}. This agreement, moreover, 
again strengthens the validity of our treatment of $\gamma^5$. Consequently
we deduce, in the notation of (2.1) and our coupling constant conventions, 
\begin{equation} 
a_4 ~=~ -~ \frac{4C_2(R)}{27} ~~~,~~~ 
a_5 ~=~ \frac{[9\zeta(3)-7]C_2(R)}{81} 
\end{equation} 

\sect{Discussion.} 

We conclude our study by remarking on possible future calculations in this 
area. The natural task to be performed next will be the $O(1/\Nf^2)$ 
corrections to the non-singlet twist-$2$ operators. Such a calculation would 
mimic the determination of the mass operator dimension but would require the 
quark dimension $\eta_2$ first. Only the abelian values are available for both 
these quantities, \cite{21}. On another front the corrections to (3.14) and 
(5.4) are needed. This would parallel the calculation of the QCD 
$\beta$-function in $1/\Nf$, \cite{41}. Both these results for the gluonic 
operators would give important insight into the $n$-dependence of the higher 
order anomalous dimensions and the $x$-behaviour of the DGLAP splitting 
functions. 

From a more mathematical physics point of view such analyses may become 
important for studying the operator content of strictly four dimensional gauge
theories which have (infrared) fixed points, \cite{42,43}. Evaluating the 
perturbative anomalous dimension of the composite operator at these points 
would be necessary to gain information on the (conformal) field content of the 
underlying theory in the perturbatively accessible region. Moreover the 
existence of fixed points such as that of Banks and Zaks in QCD, \cite{42}, for
a range of $\Nf$ values have been the subject of recent interest in 
supersymmetric theories with various gauge groups and matter content, 
\cite{43}. Therefore any information that can be determined from traditional 
field theory methods and which sum perturbation theory beyond present low 
orders such as $1/\Nf$, could be used to compare estimates of, for example, 
critical exponents deduced from exact non-perturbative arguments. 

\vspace{1cm} 
\noindent 
{\bf Acknowledgements.} The author acknowledges support for this work through
a PPARC Advanced Fellowship, thanks Dr D.J. Broadhurst for encouragement and
Drs A. Vogt and J. Bl\"umlein for useful discussion on their work. The tedious 
algebra was performed in part through use of {\sc Form}, \cite{39}, and 
{\sc Reduce}, \cite{44}. 

\appendix 
\sect{DGLAP splitting functions.} 
In this appendix we discuss the relation of our results to the DGLAP splitting
functions, $P(x,g)$. The $x$-dependence of the higher order contributions to 
these functions is currently of interest in relation to low $x$ physics. We 
recall that the anomalous dimensions of the twist-$2$ operators are related to 
the DGLAP splitting functions via a Mellin transform with the proviso that $x$,
which is the variable conjugate to the moment $n$, is restricted to the unit 
interval. First, we note that our convention is   
\begin{equation}
\int_0^1 dx \, x^{n-1} \, P (x,g) ~=~ \frac{1}{4} \gamma^{(n)}(g)  
\end{equation}
Of course (A.1) can be evaluated at the fixed point and a critical exponent
which sums the leading order in $1/\Nf$ of the splitting function can be 
deduced. So, for example, the twist-$2$ non-singlet anomalous dimension (3.5)
gives the result, 
\begin{eqnarray}  
P_{{\mbox{\footnotesize{ns}}},1}(x,g_c) &=&  
\frac{2C_2(R)(\mu-1)^2\eta^{\mbox{o}}_1}{(2\mu-1)(\mu-2)T(R)} 
\left[ \frac{(\mu^2-4\mu+1)}{(\mu-1)^2} \delta (1-x) \right. \nonumber \\  
&&~~~ \left. -~ \mu(\mu-1) x^{\mu-2} (1-x) ~+~ 
\frac{2\mu x^{\mu-2}}{(\mu-1)} ~-~ \frac{2\mu x^{\mu-2}}{(\mu-1)(1-x)_+} 
\right]  
\end{eqnarray}  
where we use the standard notation in the final term to ensure sensible 
behaviour in the $x$ $\rightarrow$ $1$ limit. We have checked that the 
$\epsilon$-expansion of (A.2) correctly reproduces the $O(1/\Nf)$ part of the 
two loop non-singlet splitting function of \cite{3}. In light of this we 
determine the three loop structure as 
\begin{equation} 
a_3^{\mbox{\footnotesize{ns}}} ~=~ \frac{17}{72} \delta(1-x) ~-~ 
\frac{(1+x^2)\ln^2 x}{18(1-x)} ~-~ \frac{(11x^2-12x+11)\ln x}{27(1-x)} ~-~ 
\frac{2(7-6x)}{27} ~+~ \frac{2}{27(1-x)_+} 
\end{equation} 

The treatment of the unpolarized and polarized singlet cases are similar but 
we detail only the latter case as the exponent has a simpler $n$-dependence. 
Moreover recent articles have examined the polarized splitting functions in the
small $x$ limit, \cite{46}. Performing the inverse Mellin transform the 
splitting function becomes 
\begin{eqnarray}  
P_{{\mbox{\footnotesize{s}}},1}^{\mbox{\footnotesize{pol}}}(x,g_c) &=&  
\frac{2C_2(R)\mu(\mu-1)^2\eta^{\mbox{o}}_1}{(2\mu-1)(\mu-2)T(R)} 
\left[ \frac{(\mu^2-4\mu+1)}{\mu(\mu-1)^2} \delta (1-x) ~-~ (\mu-1) x^{\mu-2} 
(1-x) \right. \nonumber \\
&&+~ \left. \frac{2 x^{\mu-2}}{(\mu-1)} ~-~ \frac{2x^{\mu-2}}{(\mu-1)(1-x)_+} 
{}~-~ \frac{\mu(2\mu-1) x^{\mu-2}(1-x)}{\mu(\mu-1)} \right. \\ 
&&-~ \left. \frac{(2\mu-1)(\mu-3)(\mu-4) x^{\mu-2}}{(\mu-1)}  
[x B_{1-x}(2\mu-2,1-\mu) ~-~ B_{1-x} (2\mu-2,2-\mu) ] \right] \nonumber  
\end{eqnarray}  
where $B_x(p,q)$ is the incomplete $\beta$-function which has the integral 
representation 
\begin{equation} 
B_x(p,q) ~=~ \int_0^x \, du \, u^{p-1} (1-u)^{q-1} 
\end{equation} 
We can once again perform the $\epsilon$-expansion of (A.4) and attempt to 
compare with the explicit two loop results of \cite{10,11,9}. We find from 
(A.4) 
\begin{eqnarray}  
\left[ a_1 ~-~ \frac{b_1c_1}{d_1} \right]^{\mbox{\footnotesize{pol}}} &=& -~ 
\frac{3}{2} \delta(1-x) ~-~ 6 (1+x)\ln x ~-~ \frac{(8-15x+8x^2)}{(1-x)} \\ 
\left[ a_2 ~-~ \frac{b_2c_1}{d_1} \right]^{\mbox{\footnotesize{pol}}} 
&=& \frac{1}{12}\delta(1-x) ~+~ (1+x)\ln^2 x ~+~ 10(1-x)\ln(1-x) ~+~ 
\frac{2(6-5x^2)\ln x}{3(1-x)} \nonumber \\ 
&&-~ 4 (1+x)[ \mbox{Li}_2(x) - \mbox{Li}_2(1) ] ~+~ 
\frac{2(13-21x+13x^2)}{9(1-x)} 
\end{eqnarray}  
where the dilogarithm function $\mbox{Li}_2(x)$ enters, \cite{45}. Its  
appearance, however, would seem to suggest that (A.7) does not relate to 
information in the two loop splitting matrix since $\mbox{Li}_2(x)$ is absent 
at leading order in $1/\Nf$ there, \cite{10,11}. (It does occur, for example, 
at next to leading order in $1/\Nf$ in all entries bar $\gamma^{qq}$.) The 
resolution of this rests in the triangularity property of the critical point 
mixing matrix. In order to correctly compare with the explicit perturbative 
matrix it has to be mapped to a similar structure. This is achieved by an 
invertible $2$ $\times$ $2$ matrix $R$. In other words, in matrix language,  
\begin{equation} 
P_{\mbox{\footnotesize{s}}}^{\mbox{\footnotesize{tri}}}(x,g_c) ~=~ 
R \, \left( \int_0^1 \, dx \, x^{n-1}  
P_{\mbox{\footnotesize{s}}}^{\mbox{\footnotesize{pert}}}(x,g_c) \right) R^{-1} 
\end{equation}  
As we are only interested in the $O(1/\Nf)$ $\mbox{FF}$ component this reduces 
to comparing (A.6) and (A.7) with the sum of the products of the appropriate
elements of (A.8) in $x$-space which is a combination of entries similar to the
left side of (A.6). Also since $R$ is $n$-dependent we need to express this 
product of $n$-dependent functions as a Mellin transform of a single 
$x$-dependent function for the comparison. Useful in this respect is the 
convolution formula for the Mellin transform. As splitting functions are 
defined to be zero outside the unit interval this takes the following form in 
this instance  
\begin{equation} 
{\cal M} [f_1(x)] {\cal M} [f_2(x)]  ~=~ {\cal M} \left[ \, \int_x^1 \, 
\frac{du}{u} \, f_1 \left( \frac{x}{u} \right) \, f_2 (u) \right] 
\end{equation}  
where the Mellin transform is defined to be  
\begin{equation} 
{\cal M} [f(x)] ~=~ \int_0^1 \, dx \, x^{n-1} f(x) 
\end{equation} 
So with (A.9) and, at leading order,  
\begin{eqnarray} 
a_1^{\mbox{\footnotesize{pol}}} &=& C_2(R) \left[ 1 ~+~ x ~-~ 
\frac{2}{(1-x)_+} ~-~ \frac{3}{2}\delta(1-x) \right] ~~~,~~~  
b_1^{\mbox{\footnotesize{pol}}} ~=~ -~ (2-x)C_2(R) \nonumber \\
c_1^{\mbox{\footnotesize{pol}}} &=& -~ 2(2x-1)T(R) ~~~,~~~  
d_1^{\mbox{\footnotesize{pol}}} ~=~ \frac{2T(R)}{3} \delta(1-x) \nonumber \\ 
a_2^{\mbox{\footnotesize{pol}}} &=& C_2(R)T(R) \left[  
\frac{1}{12}\delta(1-x) ~+~ \frac{(11-12x+11x^2)}{9(1-x)_+} ~+~ 
\frac{(1+x^2)}{2(1-x)}\ln x \right. \nonumber \\ 
&&-~ \left. (1-x) ~+~ (1-3x)\ln x ~+~ (1+x) \ln^2 x \right] \nonumber \\ 
b_2^{\mbox{\footnotesize{pol}}} &=& 4C_2(R)T(R) \left[ \frac{1}{9}(x+4) ~+~ 
\frac{1}{3} (2-x)\ln (1-x) \right]  
\end{eqnarray} 
it is straightforward to verify that (A.6) and (A.7) emerge. 

Finally, having established the relation of (A.4) with perturbation theory it 
is a simple exercise to produce 
\begin{eqnarray}
\left[ a_3 ~-~ \frac{b_3c_1}{d_1} \right]^{\mbox{\footnotesize{pol}}} &=& 
\frac{17}{72}\delta(1-x) \, - \, \frac{4}{3} (1+x) \ln x \ln^2(1-x) ~-~ 
\frac{10}{3} (1-x)\ln^2(1-x) \nonumber \\ 
&&-~ \frac{4}{3}\ln x \ln(1-x) ~-~ \frac{(1+x)}{9}\ln^3 x ~-~ 
\frac{(6-5x^2)}{9(1-x)}\ln^2 x \nonumber \\ 
&&+~ \frac{4}{3} (1+x)\ln x [ \mbox{Li}_2(x) - \mbox{Li}_2(1) ] ~-~ 
\frac{10}{9}(1-x)\ln(1-x) \nonumber \\ 
&& +~ \frac{2(8-9x)(2+3x)}{27(1-x)}\ln x ~+~ \frac{2}{9}(5+11x)[ \mbox{Li}_2(x)
- \mbox{Li}_2(1) ] \nonumber \\ 
&&+~ \frac{4}{3}(1+x)[ \mbox{Li}_3(x) - \mbox{Li}_3(1) - \ln x \mbox{Li}_2(x) ] 
{}~-~ \frac{4}{3}\mbox{Li}_2(1-x) \nonumber \\ 
&&+~ \frac{8}{3}(1+x)[ \mbox{Li}_3(1-x) - \ln(1-x) \mbox{Li}_2(1-x) ] ~+~ 
\frac{(55-108x+55x^2)}{27(1-x)_+} \nonumber \\ 
\end{eqnarray} 
In relation to the work of \cite{46} we deduce from (A.6), (A.7) and (A.12) 
that the leading small $x$ behaviour of these perturbative coefficients are,
respectively, 
\begin{equation} 
-~ 6 \ln x ~~~,~~~ \ln^2 x ~~~,~~~ -~ \frac{1}{9} \ln^3 x 
\end{equation} 
In respect of the large $\Nf$ and small $x$ limits, \cite{46} concludes that 
these do not commute. Although this may seem to be a disappointing result it is
worth recalling that the primary motivation of this paper is the provision of 
information on higher order coefficients of operator dimensions which can be 
compared with explicit perturbative calculations.

\newpage 
{\large 
\begin{table}[th] 
\hspace{3cm} 
{ \begin{tabular}{r||r|r} 
$n$ & $a_3 ~-~ b_3c_1/d_1$ & $[a_3 ~-~ 
b_3c_1/d_1]^{\mbox{\footnotesize{pol}}}$ \\  
\hline 
& & \\ 
$2$ & $0$ & $ - \, \frac{7}{108}$ \\
& & \\ 
$4$ & $ - \, \frac{121259}{720000}$ & $ - \, \frac{3496333}{19440000}$ \\
& & \\ 
$6$ & $ - \, \frac{3166907}{13891500}$ & $ - \, \frac{26205857}{111132000}$ \\
& & \\ 
$8$ & $ - \, \frac{1328467729}{5038848000}$  
& $ - \, \frac{466437737839}{1728324864000}$ \\
& & \\ 
$10$ & $ - \, \frac{304337312935261}{1054350180576000}$  
& $ - \, \frac{309708615382541}{1054350180576000}$ \\
& & \\ 
$12$ & $ - \, \frac{842357166098254633}{2737572318857376000}$  
& $ - \, \frac{853943993349679513}{2737572318857376000}$ \\
& & \\ 
$14$ & $ - \, \frac{42512567719680559}{131614053791220000}$  
& $ - \, \frac{343857572061363287}{1052912430329760000}$ \\
& & \\ 
$16$ & $ - \, \frac{755896148277147625515451}{2251271656795440254976000}$  
& $ - \, \frac{762758545067630140156811}{2251271656795440254976000}$ \\
& & \\ 
$18$ & $ - \, \frac{1121815282809553973842772849}
                       {3235896767484248927963904000}$  
& $ - \, \frac{1130341459796219185897273169}
                            {3235896767484248927963904000}$ \\
& & \\ 
$20$ & $ - \, \frac{78640886458671664340562623}
                       {220772683941420492161280000}$  
& $ - \, \frac{26382991363083553371777301}
                            {73590894647140164053760000}$ \\
& & \\ 
$22$ & $ - \, \frac{4248342909129791924572989157741}
                       {11650178316263341224273468364800}$  
& $ - \, \frac{4271977516708367936099843954621}
                            {11650178316263341224273468364800}$ \\
\end{tabular} } 
\end{table} }  
\vspace{1cm} 
\noindent 
{\bf Table 1. $O(1/\Nf)$ coefficients for unpolarized and polarized fermionic
twist-$2$ singlet operators at three loops as a function of moment $n$.}  

\newpage 
\epsfysize=1.5cm 
\epsfbox{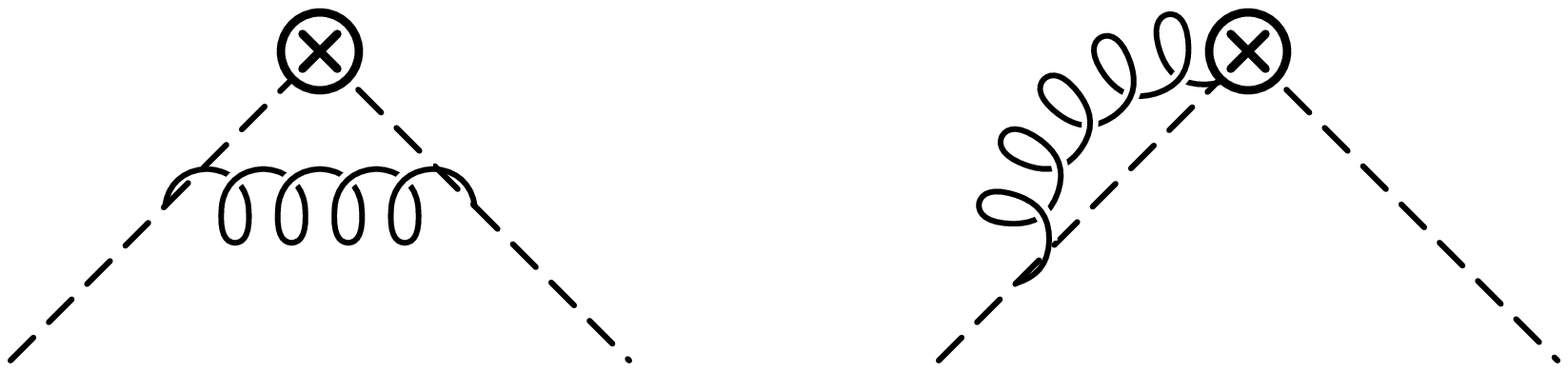} 
\vspace{1cm} 
{\bf Fig. 1. Leading order graphs for $\eta^{(n)}_{\mbox{\footnotesize{ns}}}$.} 

\vspace{2cm} 
\epsfysize=5cm 
\epsfbox{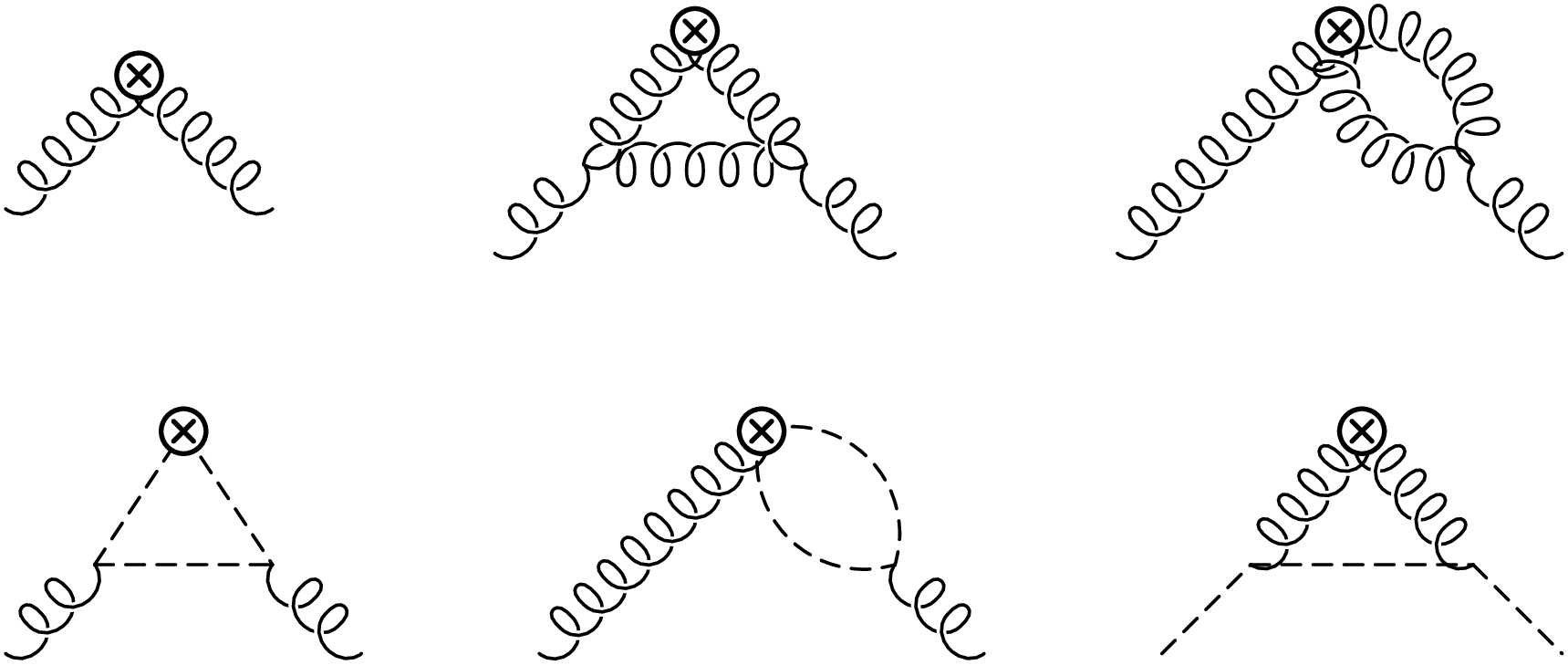} 
\vspace{1cm} 
{\bf Fig. 2. One loop graphs for singlet operators.} 

\vspace{2cm} 
\epsfysize=3cm 
\epsfbox{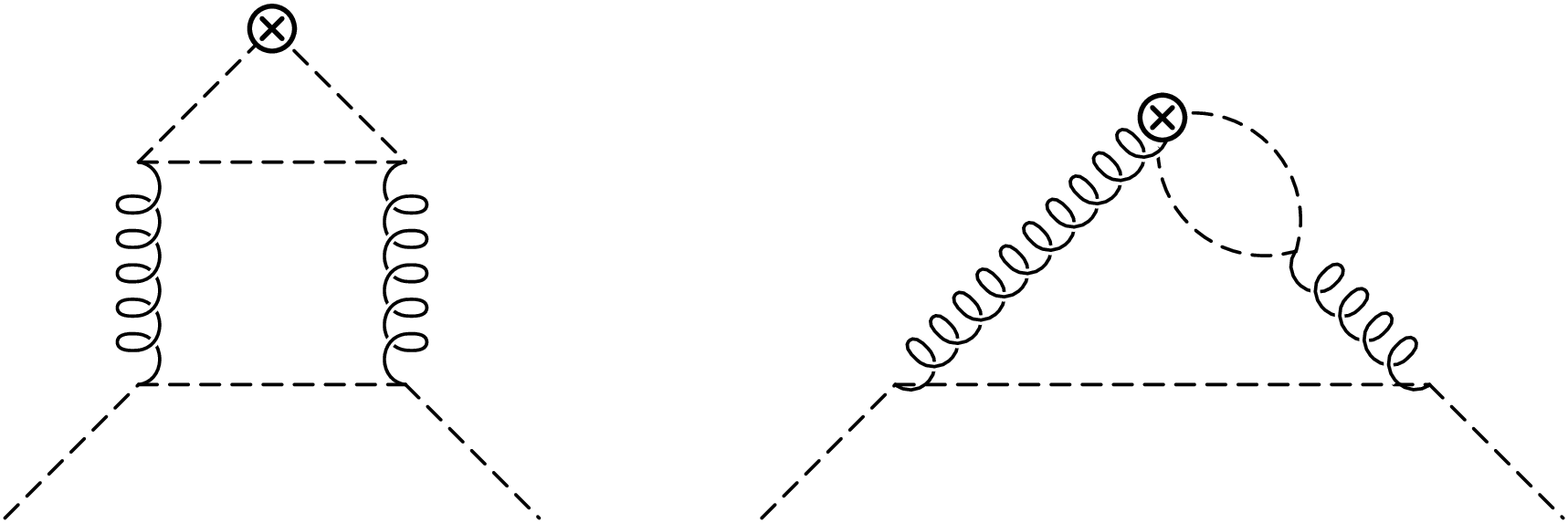} 
\vspace{1cm} 
{\bf Fig. 3. Additional graphs for singlet operators.}  

\newpage 
\vspace{2cm} 
\epsfysize=2.5cm 
\epsfbox{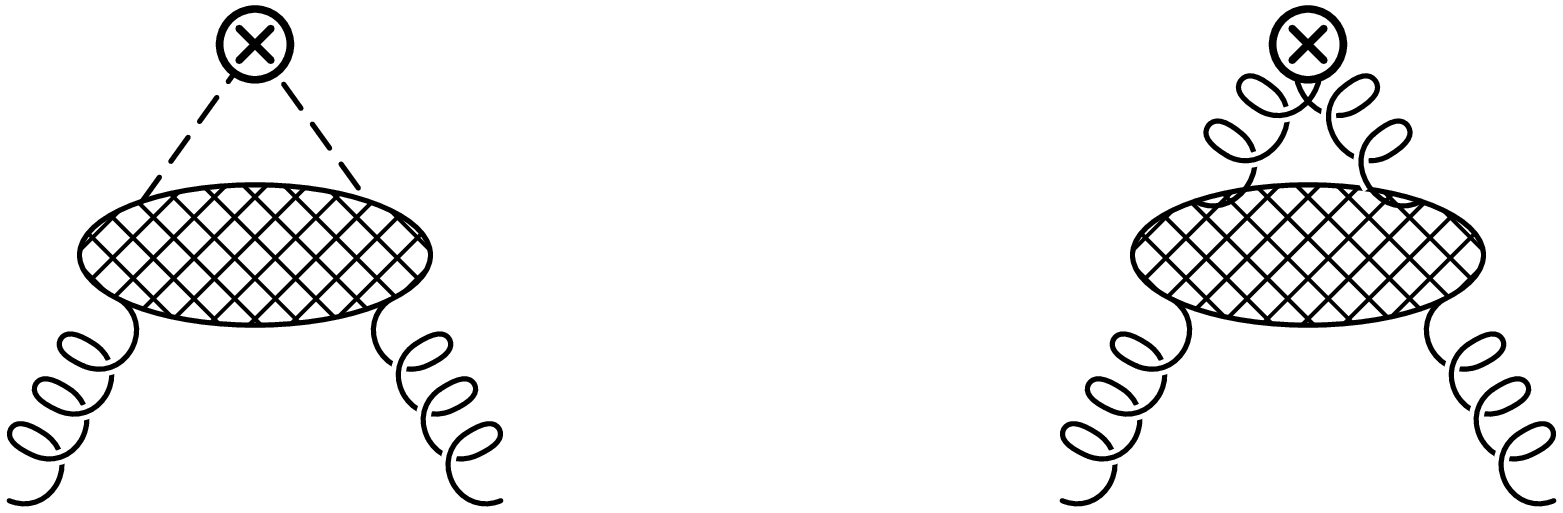} 
\vspace{1cm} 
{\bf Fig. 4. Operator insertions in gluon $2$-point functions.} 

\vspace{2cm} 
\epsfysize=1.7cm 
\epsfbox{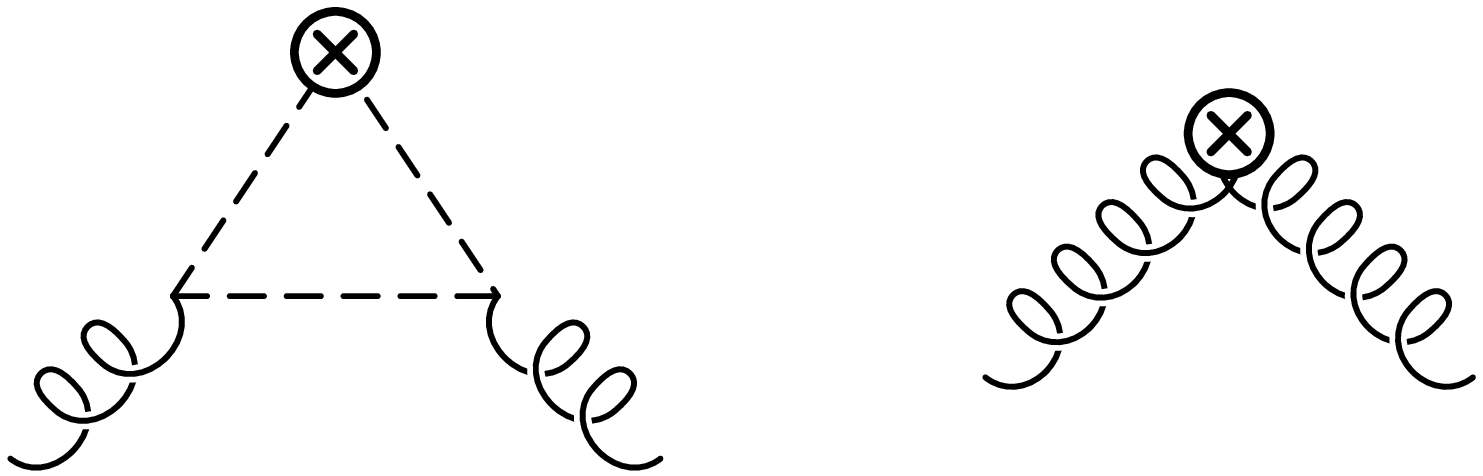} 
\vspace{1cm} 
{\bf Fig. 5. Leading order graphs for 
$\partial_\mu {\cal O}^{\mu 5}_{\mbox{\footnotesize{s}}}$ and  
$\epsilon^{\mu\nu\sigma\rho} G^a_{\mu\nu} G^a_{\sigma\rho}$ insertions.}  

\vspace{2cm} 
\epsfysize=7cm 
\epsfbox{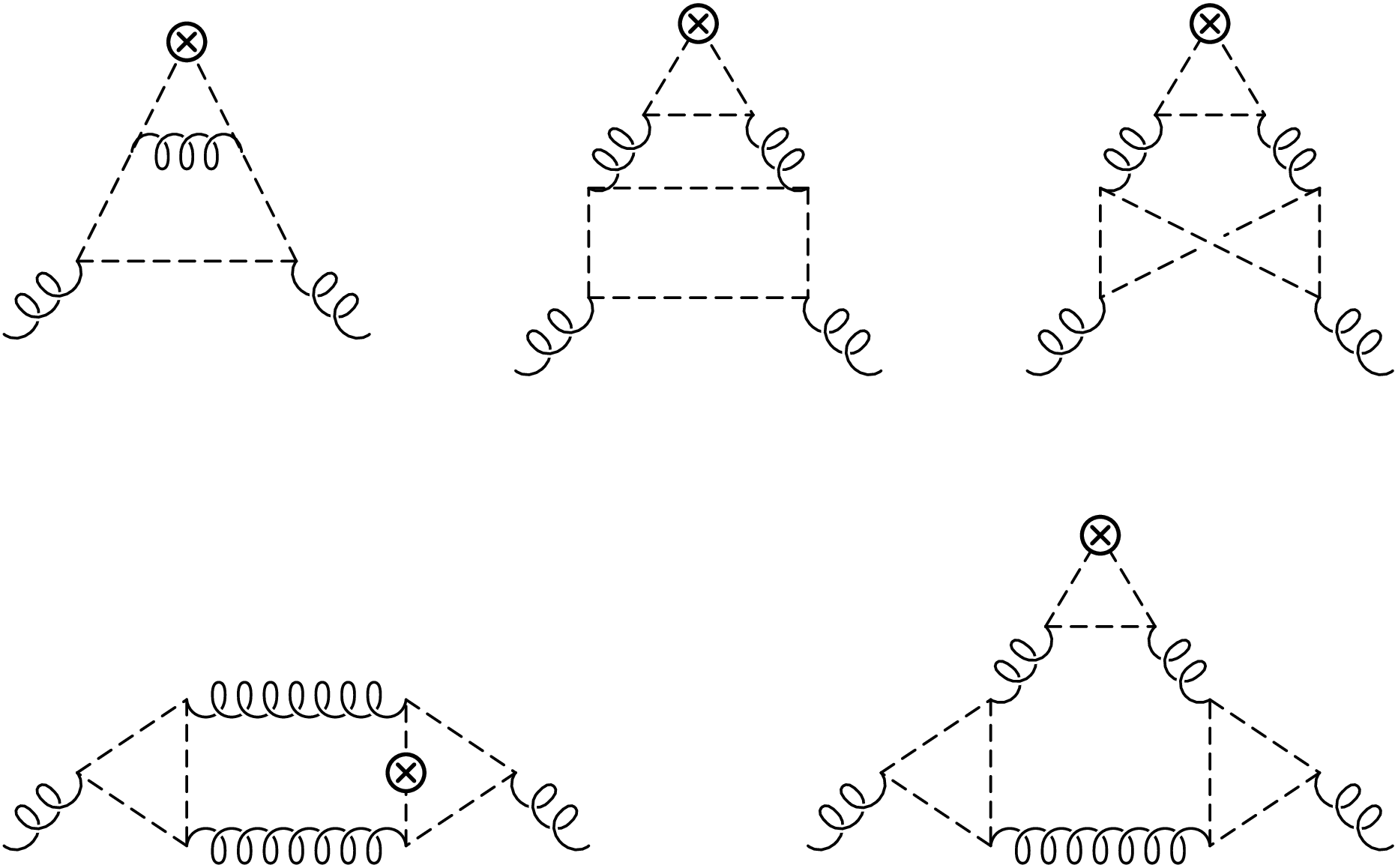} 
\vspace{1cm} 
{\bf Fig. 6. Graphs for $\partial_\mu 
{\cal O}^{\mu 5}_{\mbox{\footnotesize{s}}}$ insertion.} 

\newpage 
\vspace{2cm} 
\epsfysize=6cm 
\epsfbox{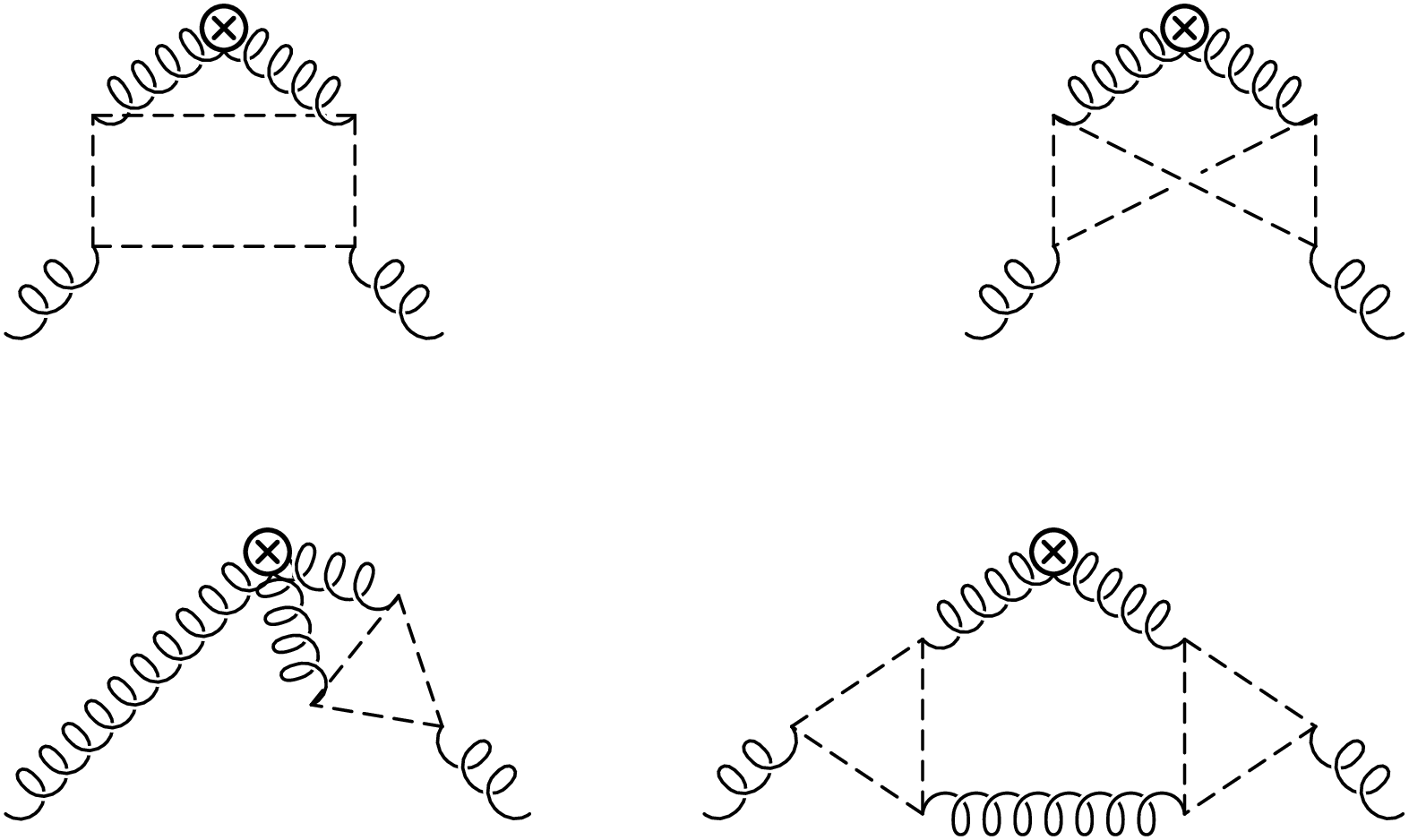} 
\vspace{1cm} 
{\bf Fig. 7. Graphs for $\epsilon^{\mu\nu\sigma\rho} G^a_{\mu\nu} 
G^a_{\sigma\rho}$ insertion.}  


\newpage 
\begin{thebibliography}{99} 
\bibitem{1} J. Ashman et al (EMC), Phys. Lett. {\bf B206} (1988), 364; Nucl. 
Phys. {\bf B328} (1989), 1; V.W. Hughes et al, Phys. Lett. {\bf B212} (1988),
511; M.J. Alguard et al (SLAC), Phys. Rev. Lett. {\bf 37} (1978), 1262; Phys.
Rev. Lett. {\bf 41} (1978), 70.  
\bibitem{2} D.J. Gross \& F.J. Wilczek, Phys. Rev. Lett. {\bf 30} (1973), 1343;
H.D. Politzer, Phys. Rev. Lett. {\bf 30} (1973), 1346.
\bibitem{3} E.G. Floratos, D.A. Ross \& C.T. Sachrajda, Nucl. Phys. {\bf
B129} (1977), 66; {\bf B139} (1978), 545(E); A. Gonz\'{a}lez-Arroyo, C. 
L\'{o}pez \& F.J. Yndur\'{a}in, Nucl. Phys. {\bf B153} (1979), 161; G. Curci, 
W. Furmanski \& R. Petronzio, Nucl. Phys. {\bf B175} (1980), 27.
\bibitem{4} E.G. Floratos, D.A. Ross \& C.T. Sachrajda, Nucl. Phys. {\bf
B152} (1979), 493.
\bibitem{5} A. Gonz\'{a}lez-Arroyo \& C. L\'{o}pez, Nucl. Phys. {\bf B166} 
(1980), 429; C. L\'{o}pez \& F.J. Yndur\'{a}in, Nucl. Phys. {\bf B183} (1981), 
157; E.G. Floratos, C. Kounnas \& R. Lacaze, Phys. Lett. {\bf B98} (1981), 
89, 285; Nucl. Phys. {\bf B192} (1981), 417.  
\bibitem{6} G. Altarelli and G. Parisi, Nucl. Phys. {\bf B126} (1977), 298; 
Yu.L. Dokshitzer, Sov. Phys. JETP {\bf 46} (1977), 641; L.N. Lipatov, Sov. J.
Nucl. Phys. {\bf 20} (1975), 95; V.N. Gribov \& L.N. Lipatov, Sov. J. Nucl. 
Phys. {\bf 15} (1972), 438.  
\bibitem{7} S.A. Larin, T. van Ritbergen and J.A.M. Vermaseren, Nucl. Phys. 
{\bf B427} (1994), 41; hep-ph/9605317. 
\bibitem{8} M.A. Ahmed \& G.G. Ross, Nucl. Phys. {\bf B111} (1976), 441. 
\bibitem{9} R. Mertig \& W.L. van Neerven, Zeit. Phys. {\bf C70} (1996), 637.  
\bibitem{10} W. Vogelsang, hep-ph/9512218. 
\bibitem{11} W. Vogelsang, hep-ph/9603366. 
\bibitem{12} T. Weigl \& W. Melnitchouk, Nucl. Phys. {\bf B465} (1996), 267. 
\bibitem{13} A.V. Manohar, ``An introduction to spin dependent deep inelastic
scattering'' Lectures Lake Louise Winter Institute, VCSD/PTH 92-10, 
hep-ph/9204207; S. Forte, ``Polarized structure functions: a theoretical 
update'', hep-ph/9511345.  
\bibitem{14} J.A. Gracey, Phys. Lett. {\bf B322} (1994), 141.
\bibitem{15} J. Zinn-Justin, ``Quantum field theory and critical phenomena''
(Clarendon Press, Oxford, 1989). 
\bibitem{16} J.A. Gracey, Phys. Lett. {\bf B318} (1993), 177.
\bibitem{17} A.N. Vasil'ev, Yu.M. Pis'mak \& J.R. Honkonen, Theor. Math. Phys.
{\bf 46} (1981), 157.
\bibitem{18} A.N. Vasil'ev, Yu.M. Pis'mak \& J.R. Honkonen, Theor. Math. Phys.
{\bf 47} (1981), 291.
\bibitem{19} A.N. Vasil'ev \& M.Yu. Nalimov, Theor. Math. Phys. {\bf 55}
(1982), 163; {\bf 56} (1983), 15;  A.N. Vasil'ev, N.Yu. Nalimov \& J.R. 
Honkonen, Theor. Math. Phys. {\bf 58} (1984), 111.
\bibitem{20} S.E. Derkachov, N.A. Kivel, A.S. Stepanenko \& A.N. Vasil'ev, 
Theor. Math. Phys. {\bf 92} (1992), 486; Theor. Math. Phys. {\bf 94} (1993), 
127; A.N. Vasil'ev \& A.S. Stepanenko, Theor. Math. Phys. {\bf 97} (1993), 
1349.  
\bibitem{21} J.A. Gracey, Phys. Lett. {\bf B317} (1993), 415; Int. J. Mod. 
Phys. {\bf A9} (1994), 567, 727; Nucl. Phys. {\bf B414} (1994), 614. 
\bibitem{22} A. Hasenfratz \& P. Hasenfratz, Phys. Lett. {\bf B297} (1992), 
166. 
\bibitem{23} W.E. Caswell, Phys. Rev. Lett. {\bf 33} (1974), 244; D.R.T. Jones,
Nucl. Phys. {\bf B75} (1974), 531; E.S. Egorian \& O.V. Tarasov, Teor. Mat. 
Fiz. {\bf 41} (1979), 26.  
\bibitem{24} O.V. Tarasov, A.A. Vladimirov \& A.Yu. Zharkov, Phys. Lett. 
{\bf 93B} (1980), 429; S.A. Larin \& J.A.M. Vermaseren, Phys. Lett. {\bf B303} 
(1993), 334.
\bibitem{25} H. Georgi \& H.D. Politzer, Phys. Rev. {\bf D8} (1974), 416; D.J. 
Gross \& F.J. Wilczek, Phys. Rev. {\bf D8} (1973) 3633.
\bibitem{26} G. 't Hooft \& M. Veltman, Nucl. Phys. {\bf B44} (1972), 189. 
\bibitem{27} D.A. Akyeampong \& R. Delbourgo, Nuovo Cim. {\bf 17A} (1973), 
578; M. Chanowitz, M. Furmam \& I. Hinchliffe, Nucl. Phys. {\bf B159} (1979), 
225. 
\bibitem{28} P. Breitenlohner \& D. Maison, Comm. Math. Phys. {\bf 52} (1977),
11. 
\bibitem{29} P. Baikov \& V.A. Il'in, Theor. Math. Phys. {\bf 88} (1991), 789.
\bibitem{30} S.A. Larin, Phys. Lett. {\bf B303} (1993), 113. 
\bibitem{31} T.L. Trueman, Phys. Lett. {\bf B88} (1979), 331. 
\bibitem{32} J. Kodaira, Nucl. Phys. {\bf B165} (1980), 129. 
\bibitem{33} D.J. Broadhurst \& A.L. Kataev, Phys. Lett. {\bf B315} (1993), 
179.
\bibitem{34} S.L. Adler, Phys. Rev. {\bf 177} (1969), 2426. 
\bibitem{35} J.S. Bell \& R. Jackiw, Nuovo Cim. {\bf 60A} (1969), 47. 
\bibitem{36} A.L. Adler \& W. Bardeen, Phys. Rev. {\bf 182} (1969), 1517. 
\bibitem{37} A.A. Ansel'm \& A.A. Johansen, Pis'ma v ZhETF {\bf 49} (1989), 
185; ZhETF {\bf 96} (1989), 1181. 
\bibitem{38} D. Espriu \& R. Tarrach, Zeit. Phys. {\bf C16} (1982), 77. 
\bibitem{39} M. Bos, Nucl. Phys. {\bf 404} (1993), 215. 
\bibitem{40} J.A.M. Vermaseren, ``{\sc Form}'' version 2.2c, CAN publication,  
(1992). 
\bibitem{41} J.A. Gracey, Phys. Lett. {\bf B373} (1996), 178. 
\bibitem{42} T. Banks \& A. Zaks, Nucl. Phys. {\bf B196} (1982), 189. 
\bibitem{43} N. Seiberg \& E. Witten, Nucl. Phys. {\bf B426} (1994), 19; N. 
Seiberg, Nucl. Phys. {\bf B435} (1995), 129. 
\bibitem{44} A.C. Hearn, ``{\sc Reduce} Users Manual'' version 3.4, Rand 
publication CP78, (1991). 
\bibitem{45} L. Lewin, ``Dilogarithms and associated functions'' (Macdonald, 
London, 1958).  
\bibitem{46} J. Bl\"{u}mlein \& A. Vogt, Phys. Lett. {\bf B370} (1996), 149; 
hep-ph/9606254. 
\end{thebibliography}
\end{document}